\documentclass[fleqn,usenatbib]{mnras}

\usepackage{newtxtext,newtxmath}
\usepackage{comment}
\usepackage{xspace}

\usepackage[T1]{fontenc}

\DeclareRobustCommand{\VAN}[3]{#2}
\let\VANthebibliography\thebibliography
\def\thebibliography{\DeclareRobustCommand{\VAN}[3]{##3}\VANthebibliography}

\usepackage{graphicx}	
\usepackage{amsmath}	
\usepackage{threeparttable} 

\begingroup
\catcode`\_=\active
\gdef_#1{\ensuremath{\sb{\mathrm{#1}}}}
\endgroup
\mathcode`\_=\string"8000
\catcode`\_=12

\def\hide#1{}
\newcommand{\cc}{cm$^{-3}$}                          
\newcommand{\hcc}{H~cm$^{-3}$}                       
\newcommand{\hm}{H$_2$}                              
\newcommand{\kms}{{\rm km}~{\rm s}^{-1}}             
\newcommand{\msun}{{\rm M}_{\odot}}                  
\newcommand{\msyr}{{\rm M}_{\odot}~{\rm yr}^{-1}}    
\newcommand{\mskyr}{{\rm M}_{\odot}~{\rm kyr}^{-1}}  
\newcommand{\nh}{n_{H}}                              
\newcommand{\nchm}{N_{\mbox{\scriptsize{H$_2$}}}}    
\newcommand{\nhm}{n_{\mbox{\scriptsize{H$_2$}}}}     
\newcommand{\ramses}{{\sc ramses}}                
\newcommand{\ramsesrt}{{\sc ramses-rt}}           

\newcommand{\HI}{\mbox{H\,{\sc i}}\xspace}
\newcommand{\HII}{H\,{\sc ii}\xspace}
\newcommand{\HeI}{\mbox{He\,{\sc i}}\xspace}
\newcommand{\HeII}{He\,{\sc ii}\xspace}

\title[First stars in an X-ray background]{Population~III star formation in an X-ray background: III.\\ Periodic radiative feedback and luminosity induced by elliptical orbits}

\author[J. Park, M. Ricotti and K. Sugimura]{
Jongwon Park,$^{1}$
Massimo Ricotti,$^{1}$
and Kazuyuki Sugimura$^{2,3}$
\\
$^{1}$Department of Astronomy, University of Maryland, College Park, MD 20742, USA\\
$^{2}$The Hakubi Center for Advanced Research, Kyoto University, Sakyo, Kyoto 606-8501, Japan\\
$^{3}$Department of Physics, Graduate School of Science, Kyoto University, Sakyo, Kyoto 606-8502, Japan\\
}

\date{Accepted XXX. Received YYY; in original form ZZZ}

\pubyear{2022}

\begin{document}
\label{firstpage}
\pagerange{\pageref{firstpage}--\pageref{lastpage}}
\maketitle

\begin{abstract}
We model Pop~III star formation in different FUV and X-ray backgrounds, including radiation feedback from protostars. We confirm previous results that a moderate X-ray background increases the number of Pop~III systems per unit cosmological volume, but masses and multiplicities of the system are reduced. The stellar mass function also agrees with previous results, and we confirm the outward migration of the stars within the protostellar discs. We find that nearly all Pop~III star systems are hierarchical, i.e., binaries of binaries. Typically, two equal-mass stars form near the centre of the protostellar disc and migrate outward. Around these stars, mini-discs fragment forming binaries that also migrate outward. Stars may also form at Lagrange points L4/L5 of the system. Afterward, star formation becomes more stochastic due to the large multiplicity, and zero-metallicity low-mass stars can form when rapidly ejected from the disc. Stars in the disc often have eccentric orbits, leading to a periodic modulation of their accretion rates and luminosities. At the pericenter, due to strong accretion, the star can enter a red-supergiant phase reaching nearly Eddington luminosity in the optical bands ($m_{AB}\sim 34$ for a 100~M$_\odot$ star at z=6). During this phase, the star, rather than its nebular lines, can be observed directly by {JWST}, if sufficiently magnified by a gravitational lens. The $\sim10,000$~AU separations and high eccentricities of many Pop~III star binaries in our simulations are favorable parameters for IMBH mergers -- and gravitational waves emission -- through orbital excitation by field stars.
\end{abstract}


\section{Introduction}
The James Webb Space Telescope ({\it JWST}) and next-generation observatories, such as Nancy Grace Roman Space Telescope ({\it NRST}), will enable us to observe and study the properties of high-$z$ galaxies and perhaps discover the first metal-free stars (or Pop~III stars). The properties of the first generation of galaxies are sensitive to the poorly constrained models of Pop~III star formation because the first stars enrich and spread in the intergalactic medium the first heavy elements (\citealp*{RicottiGS:2002a, RicottiGS:2002b}; \citealp{greif2010,wise2012}; \citealp*{ Safranek-Shrader2014}; \citealp*{chiaki2018}; \citealp{Abe2021}) that enable and regulate the formation of the second generation stars.

One of the key properties of Pop~III stars is their initial mass function (IMF). The mass determines the luminosity of the stars \citep{schaerer2002}, therefore the strength of their UV feedback (FB) which blows out the gas in the host haloes \citep{wise2008}, but also their metal yield and the SN/hypernova explosion energy \citep[e.g.,][]{nomoto2006}. For instance, Pop~III stars are thought to explode as pair-instability supernovae (PISNe) if their masses are $140~\msun < M <260~\msun$ \citep{heger2002}. The death of these massive stars as hypernovae/PISNe is detectable by {\it JWST} or {\it NRST} \citep{whalen2014}, thereby providing constraints on the number and typical masses of Pop~III stars.

Another poorly known but important aspect of Pop~III stars is their multiplicity. The formation of multiple Pop~III protostars systems has been observed in high-resolution simulations (\citealp*{stacy2010}; \citealp{clark2011,stacy2013,sugimura2020}), offering a channel of formation of intermediate-mass binary black holes (BBHs). Given the large remnant masses of Pop~III stars, the primordial origin of mergers of BBHs detected by VIRGO and LIGO collaboration \citep{abbott2016,abbott2017} is an open possibility, although the expected merger rates are small and uncertain \citep{hartwig2016}. Pop~III binaries can also evolve to high-mass X-ray binaries (HMXBs) which emit X-ray photons. Owing to their large mean free paths, X-rays produce a background radiation that can efficiently heat and partially ionize the intergalactic medium and thus have important effects on galaxy formation and reionization (\citealp{RicottiO:2004}; \citealp*{RicottiOG:2005}; \citealp{mirabel2011,xu2014,jeon2014a}). Whether HMXBs are a dominant source of X-rays in the early universe depends on the number of Pop~III binaries \citep{stacy2013,hummel2015} and therefore the properties of primordial protostellar discs leading to their fragmentation and the formation of multiple Pop~III star systems.
Prediction of the typical masses and multiplicity of Pop~III stars requires modelling several physical processes. One of them is the unknown intensity of the X-ray radiation background that must be present because of the unavoidable emission by Pop~III supernovae, but also HMXBs and accreting intermediate-mass black holes \citep{ParkR:2011}. X-ray photons boost molecular hydrogen (\hm) formation in the gas phase by partially ionizing the intergalactic medium as shown by various authors (\citealp*{haiman2000,RicottiGS:2001, venkatesan2001}; \citealp{ricotti2016}). These authors claimed that an X-ray background may compensate the negative FB of \hm-dissociating photons in Lyman-Werner (LW) bands. The impact of X-ray irradiation on the typical masses and multiplicity of Pop~III stars was first studied in the seminal work of \citet{hummel2015}. However, their conclusion was that X-rays play a minor role in Pop~III star formation. On the other hand, in \citet*{park2021a} and \citet*[][Paper~I and Paper~II hereafter]{park2021b}, using a set of hydrodynamics simulations we showed that an X-ray background can increase the number of minihalos forming Pop~III stars but at the same time it reduces their typical masses and multiplicity.
\begin{figure*}
    \centering
	\includegraphics[width=0.95\textwidth]{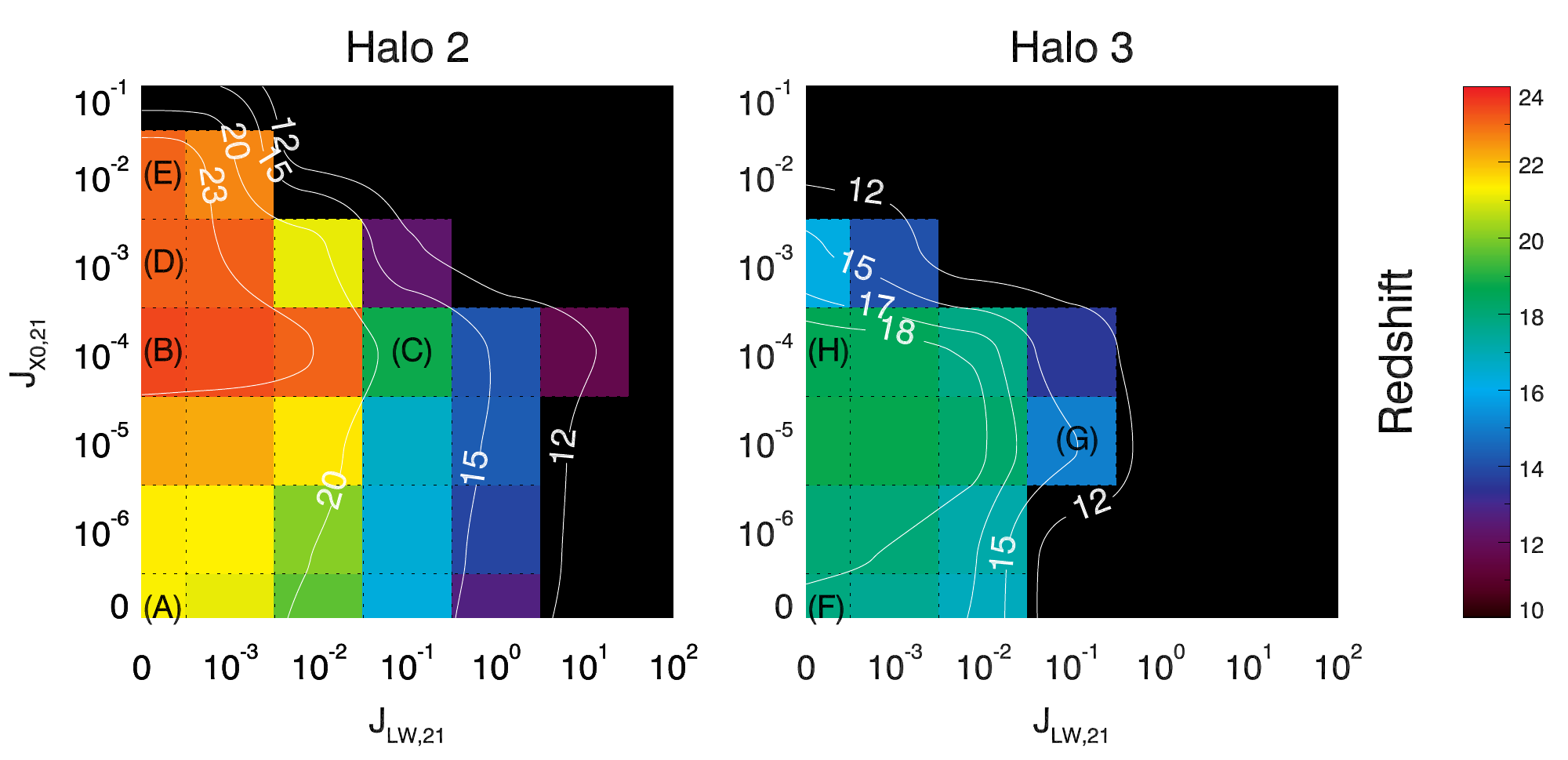}
    \caption{Redshift of collapse to a hydrogen number density $\nh = 10^7$~\hcc\ as a function of LW and X-ray background intensities. The left and right panels are the results of Halo~2 ($M_{dm}\sim 5\times 10^6~\msun$ at $z=15$) and Halo~3 ($M_{dm}\sim 7\times 10^5~\msun$ at $z=15$), respectively. The white lines show the interpolated contour plot of constant redshifts of collapse. The labels (A to H) refer to the simulations in Table~\ref{tab:sim}, see the text.}
    \label{fig:redshift}
\end{figure*}

An important physical process affecting the final masses of Pop~III stars is the FUV and EUV radiative feedback (hereafter, RFB\footnote{Both X-ray/LW background radiation and protostellar radiation effects are called `feedback', but we term only the latter 'RFB'.}) from the stars themselves. Protostars emit FUV photons in the LW bands that can dissociate \hm\ in the protostellar disc, and hydrogen ionizing EUV photons that erode and evaporate circumstellar discs, eventually terminating the gas accretion onto them, hence determining their final masses \citep[{e.g.},][]{mckee2008,hosokawa2011}. The impact of RFB on Pop~III star formation was studied in various minihaloes by several authors \citep{hirano2014,hirano2015,hosokawa2016, sugimura2020}. In Paper~I and II we studied the effect of X-rays on Pop~III star formation but we did not directly simulate RFB by UV radiation from protostars, instead we used an empirical relationship by \citet[][hereafter HR15]{hirano2015} to estimate the effect of UV FB and the final masses of Pop~III stars. In this work, we build on the results in Paper~I and II by simulating the combined effect of RFB and an external X-ray background. This is the first study investigating the masses and multiplicity of Pop~III stars regulated by RFB in different X-ray and LW radiation backgrounds.

A physical process that is potentially important in the presence of a strong X-ray background is gas cooling by hydrogen-deuteride (HD) molecules. Thanks to its small dipole moment HD is an efficient coolant in the early universe (\citealp*{lipovka2005}; \citealp{galli2013}) and enables the primordial gas to cool down to a few tens of Kelvins. Cooling by HD, therefore, can lower the mass of Pop~III stars down to a few $\sim 10~\msun$ \citep*{nagakura2005}. This effect is thought to be important in the formation of second-generation Pop~III stars rather than the first-generation.\footnote{Sometimes they are called Pop~III.2 and Pop~III.1 in the literature, but we do not make a distinction between them in this work.} For instance, Pop~III stars forming in the gas previously ionized by other Pop~III stars (recombining \HII regions, see \cite{RicottiGS:2001}) have lower masses due to the enhanced HD formation and cooling \citep{yoshida2007}. Cosmic rays \citep{nakauchi2014} and/or an X-ray background \citep{jeon2014a,hummel2015}, can also boost the abundance of HD to allow significant cooling to low temperatures. 

In this work we explore the formation and evolution of Pop~III protostars in primordial discs regulated by RFB from protostars, also exploring the effects of HD cooling, irradiated by different intensities of an external LW and X-ray backgrounds. This includes their mass growth and motions in the discs. We simulate, with higher resolution with respect to our previous work, a subset of the grid of simulations presented in Paper~I and II, either including or excluding RFB from the accreting protostars. We run the simulations for about 100~kyr after the formation of the protostar, a time comparable to the accretion timescale of the protostars (HR15).

The rest of the paper is structured as follows. In Section~\ref{sec:sim} we introduce our simulations and methods with emphasis on the improvements with respect to Paper~I and Paper~II. In Section~\ref{sec:results}, we present the results of the simulations and discuss RFB effect in each. In Section~\ref{sec:HD}, we discuss the effects of including HD cooling. In Section~\ref{sec:discussion}, we discuss the implication of our results. In Section~\ref{sec:sum} we summarize the paper and present the key results.

\section{Simulations}
\label{sec:sim}
The simulations in this paper are run with the adaptive mesh refinement (AMR) radiative hydrodynamics (RHD) code \ramsesrt\ \citep{teyssier2002,rosdahl2013}. \ramsesrt\ has been used in RHD simulations of galaxy formation \citep[e.g.,][]{kimm2017,katz2017} and metal-rich star formation (\citealp*[e.g.,][]{he2019,he2020}; \citealp{commercon2022}; \citealp{han2022}). To our best knowledge, our simulations are the first ones carried out with \ramsesrt\ to model RFB from Pop~III protostars. For this reason, in this section we describe in detail the relevant physics modules we added to the code to simulate Pop~III star formation.

\subsection{Grid of cosmological initial conditions and zoom-in simulations}
\label{sec:grid}

Pop~III star formation occurs in a protostellar disc and therefore the disc must be resolved with a high spatial resolution ($\sim$~AU) in the simulation. Therefore, to resolve the disc in a cosmological simulation we need to cover a prohibitively large dynamical range (from Mpc to AU). To circumvent this problem we use the same approach taken in several previous studies (\citealp{smith2011,hirano2014}; HR15; \citealp{sugimura2020}). First, we run cosmological zoom-in simulations as in Paper~I and Paper~II. These include simulations of three haloes of different masses irradiated by various LW and soft X-ray background intensities (see Paper~I for details of the setup). We also include some previously neglected physics (HD cooling and X-ray self-shielding) described below. We stop the zoom-in simulations when the central hydrogen number density reaches $\nh \sim 10^7$~\hcc\ or at $z \sim 10$. Fig.~\ref{fig:redshift} shows the redshift when the gas density in the haloes reaches $\nh \sim 10^7$~\hcc\ for a grid of simulations with different intensities of the LW background (x-axis) and X-ray background (y-axis) for two haloes. Within this set of low-resolution simulations we pick a few representative cases to continue running further following the formation and growth of the protostars under the effect of RFB. We choose five simulations for Halo~2 (labelled from A to E in the left panel of Fig.~\ref{fig:redshift}) and three from Halo~3 (F, G and H, right panel). From A to E have increasing X-ray intensities in general but B and C have the same with different LW intensities. D and E belong to the strong X-ray case (J$_{X0,21} \geq 10^{-3}$) in Paper~I and Paper~II. F, G and H from Halo~3 have zero and intermediate X-ray intensities (J$_{X0,21} = 0, 10^{-5}$, and $10^{-4}$, respectively). We do not show the results for the strong X-ray case in Halo~3 (J$_{X0,21} = 10^{-3}$) because accidentally this halo experienced a recent merger near the time of Pop~III stars formation. Therefore, the star formation properties do not follow the expected trends with X-ray intensity. Studying the effects of a major merger is beyond the scope of this paper and will be studied in future work. To see potential LW background effects, we select two simulations with a non-zero LW background (C and G). The name of each selection is shown in Table~\ref{tab:sim}. With the same initial condition we run two sets of simulations: one including RFB and one without RFB. We find that the growth of stars in the simulations with an X-ray background but without LW background (B, D, E, and H) stops rather quickly and therefore we cover a relatively short time evolution. For E, however, we evolve the simulation for a longer time (115~kyr) to confirm that indeed stars do not resume growing at later times.
\begin{table*}
    \caption{Summary of simulations.}
    \footnotesize
    \begin{threeparttable}
    	\centering
    	\label{tab:sim}
    	\begin{tabular}{ | l | c | c | c | c | c | c | c | }
		    \hline
            Label\tnote{a} & Halo & RFB & J$_{X0,21}$ & J$_{LW,21}$ & $t_{final}$ [kyr] & $M_{final}$ [$\msun$]& Multiplicity, $N_{final}$  \\
		    \hline
            Run~A(\_noFB)\tnote{b} & Halo~2 & Yes~(No) & $0$ & $0$ & $68~(81)$ & $557~(1086)$ & 4 (7) \\
            Run~B(\_noFB) & Halo~2 & Yes~(No) & $10^{-4}$ & $0$ & $42~(71)$ & $183~(689)$ & 3 (6)  \\
            Run~C(\_noFB) & Halo~2 & Yes~(No) & $10^{-4}$ & $10^{-1}$ & $92~(125)$ & $579~(972)$ & 4 (5) \\
            Run~D(\_noFB) & Halo~2 & Yes~(No) & $10^{-3}$ & $0$ & $67~(115)$ & $156~(281)$ & 2 (2)  \\
            Run~E(\_noFB)\tnote{b} & Halo~2 & Yes~(No) & $10^{-2}$ & $0$ & $115~(185)$ & $105~(433)$ & 2 (5)  \\
            Run~E\_noHD & Halo~2 & Yes & $10^{-2}$ & $0$ & $18$ & $95$ & 2  \\
            \hline
            Run~F(\_noFB)\tnote{b} & Halo~3 & Yes~(No) & $0$ & $0$ & $108~(87)$ & $338~(470)$ & 7 (5)   \\
            Run~G(\_noFB)\tnote{b} & Halo~3 & Yes~(No) & $10^{-5}$ & $10^{-1}$ & $142~(148)$ & $539~(801)$ & 10 (10)   \\
            Run~H(\_noFB) & Halo~3 & Yes~(No) & $10^{-4}$ & $0$ & $33~(90)$ & $203~(603)$ & 2 (4)  \\
            \hline
	    \end{tabular}
	    \begin{tablenotes}
	        \item[a] Following the labels in Fig.~\ref{fig:redshift}.
                \item[b] Movies of the fiducial simulations are available in the supplementary material.
        \end{tablenotes}
    \end{threeparttable}
\end{table*}

After selecting the simulations we extract the central 2~pc of these runs to generate new initial conditions and perform simulations in non-cosmological settings. Throughout the paper, we call these new simulations ``SF (star formation) simulations.'' We do not include dark matter particles because their density in the halo core is several orders of magnitude lower than the gas density. We confirmed that the SF simulations produce results that agree with simply running the zoom-in cosmological simulations further without extracting a 2~pc box region. Even though the simulations are not identical due to the chaotic nature of the fragmentation process, the results are qualitatively identical and the SF simulations run more efficiently than the zoom-in simulations. In the SF simulations the base AMR level is 7 and cells are refined with Jeans criterion  \citep{truelove1997} $N_{J} = 16$ up to AMR level 15. The cell size at the maximum AMR level is $\Delta x = 2~{\rm pc}/2^{15} = 6.10 \times 10^{-5}~{\rm pc} = 12.6~{\rm AU}$. Since we aim at running multiple simulations with different radiation backgrounds for about $\sim 50-100$~kyr from the formation of the first protostar, we adopt a spatial resolution a few times lower than other recent simulations of Pop~III star formation \citep[$\sim 3$~AU,][]{hosokawa2016,sugimura2020}, but significantly higher than that of our previous works ($\sim 60$~AU, Paper~I and Paper~II). 

\begin{figure}
    \centering
	\includegraphics[width=0.48\textwidth]{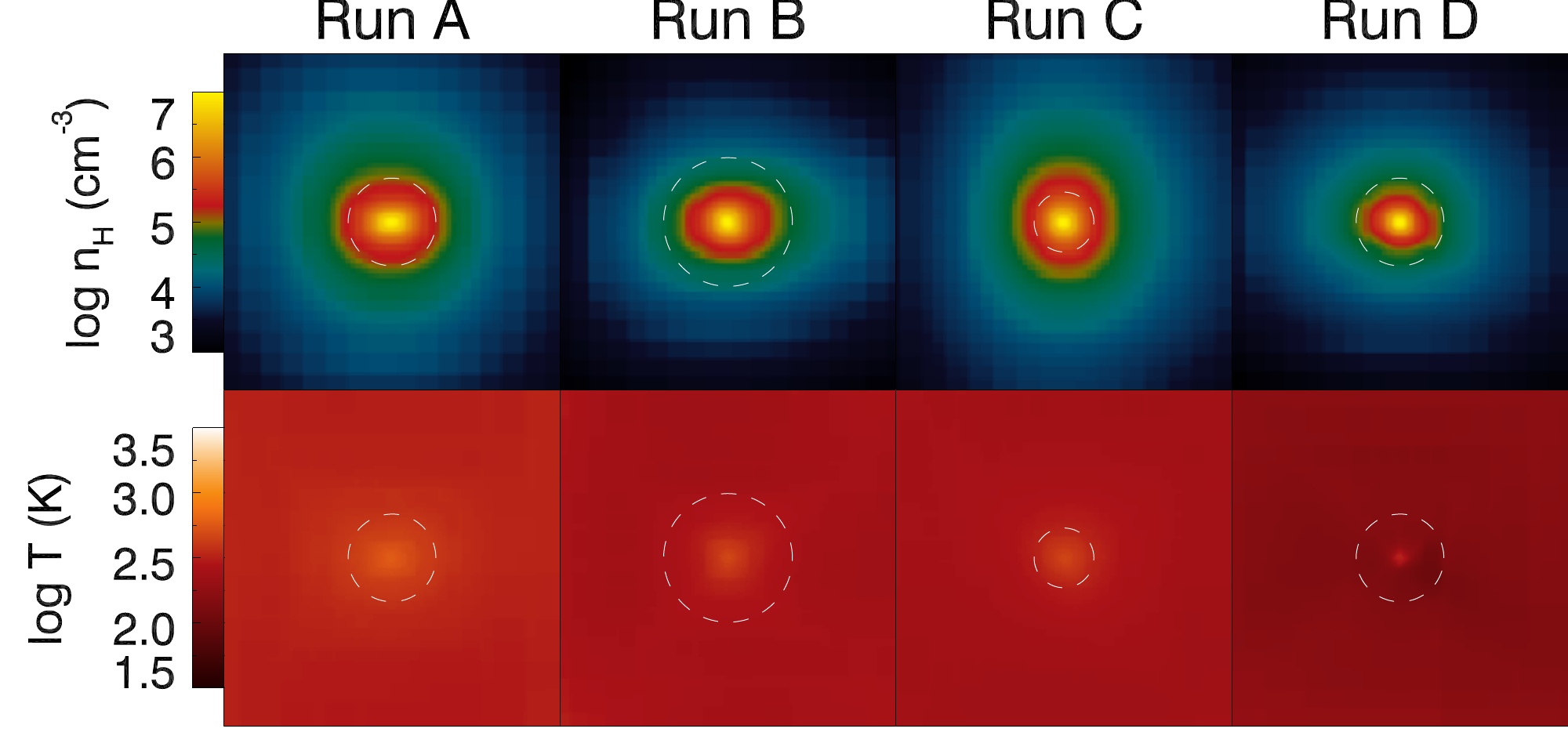}
	\includegraphics[width=0.48\textwidth]{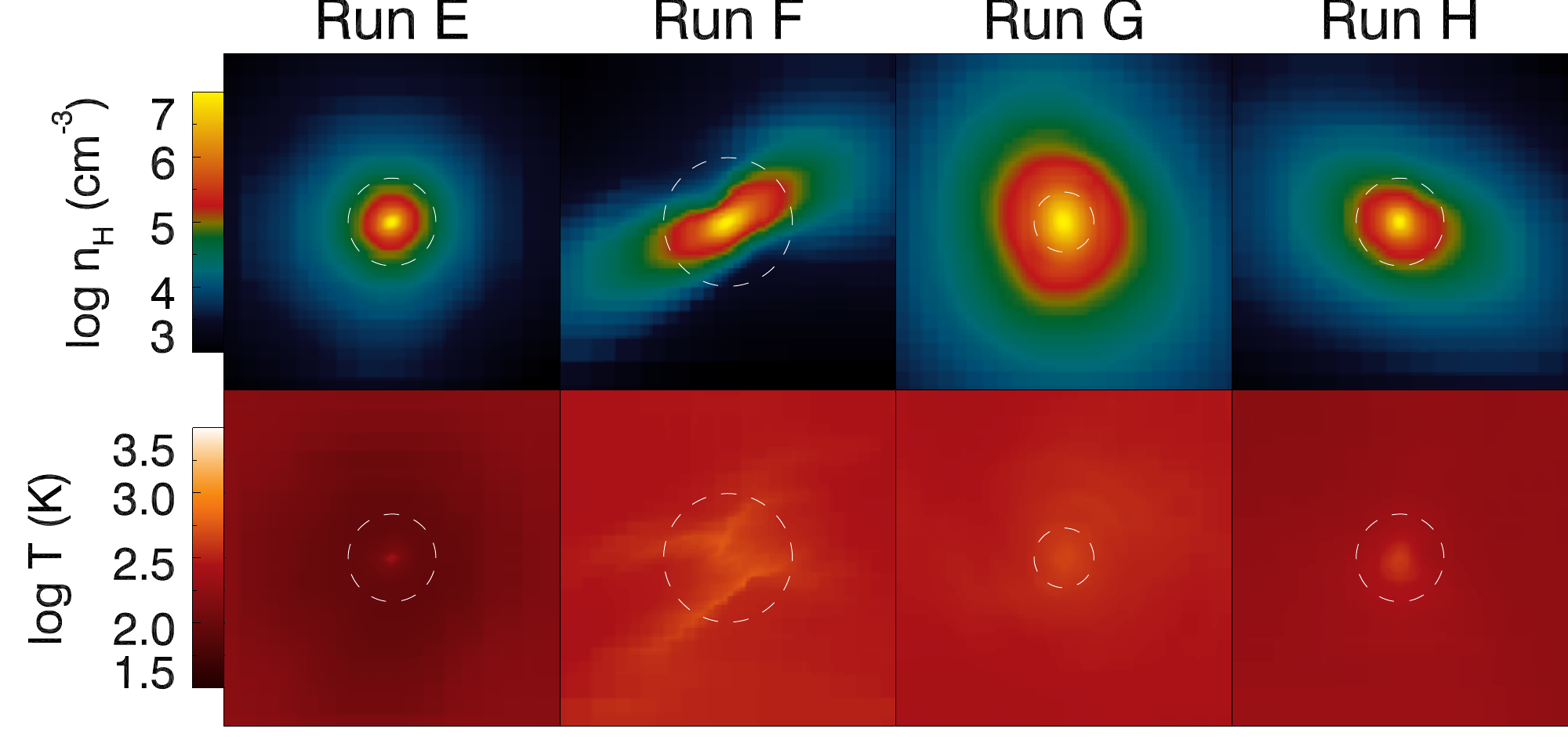}
    \caption{Hydrogen number density (top) and temperature (bottom) slices of the initial conditions for the simulations in Table~\ref{tab:sim}. White dashed circles indicate the effective radius of the collapsing cloud  (\citealp{hirano2014}; HR15). The scale of each image is 2~pc on each side.}
    \label{fig:ic}
\end{figure}

The initial conditions are shown in Fig.~\ref{fig:ic}. The characteristic radius of the collapsing cloud (\citealp{hirano2014}; HR15) does not exceed 1~pc (dashed lines). Initial conditions of Halo~2 display a clear trend with the X-ray intensity with the gas cloud tending to be colder ($\sim 100~$K) and smaller in size in X-rays. When irradiated by LW, on the other hand, the gas cloud becomes larger in size due to the delayed Pop~III star formation and therefore a larger halo mass at a lower redshift.

\subsection{Primordial \hm\ and HD chemistry and cooling}

We refer to Paper~I for a description of the primordial gas chemistry and cooling function of \hm. In this work we also include the effect of HD cooling that can become important in cases with strong X-ray irradiation due to the significantly increased \hm\ molecular fraction and low gas temperature.
To consider the HD cooling we implement 6 reactions including D, D$^+$, and HD following Table~2 of \citet*{nakauchi2019} and the HD cooling function from \citet{lipovka2005}. The HD cooling is only included in cosmological zoom-in simulations and ignored in SF runs because it is a sub-dominant process at higher densities \citep{omukai2005}.

\subsection{Self-shielding of Background Radiation}
\label{sec:sshd}
Although X-ray photons have small ionization cross-sections, they are shielded by the dense central gas in a halo. This X-ray shielding may shut down the additional ionization/heating in the central regions of the protostellar core, weakening the FB effect of X-rays as found in \cite{hummel2015}. In Paper~II we also found a weakening effect of self-shielding for strong X-ray irradiation cases, but the self-shielding does not make a meaningful difference in weak or moderate X-ray backgrounds (see Appendix~A in Paper~II). This is because Pop~III properties are determined when the central density reaches moderate densities ($\sim 10^4$~\hcc), before self-shielding becomes important. However, we include this effect anyway in this work for completeness. For the X-ray shielding, the ionization and heating rate by X-ray background is multiplied by the self-shielding factor ($0 < f_{shd} < 1$):
\begin{equation}
    f_{shd} = ~\exp({-N_{c} \overline{\sigma}}),
\end{equation}
where $N_{c}$ is the column density and $\overline{\sigma}$ is the averaged ionization cross-section in the X-ray range $[200~{\rm eV}, \infty)$. We assume the halo is spherically symmetric and thus the column density is only a function of the distance, $r$, from the centre:
\begin{equation}
    N_{c}(r) = \int_{R_{vir}}^{r} n (r') |{\rm d}r'|
\end{equation}
where $R_{vir}$ is the virial radius of the halo and $n(r')$ is the number density profile. As a side note, \citet{hummel2015} considered the geometry of the protostellar disc in the self-shielding factor. We use the above simple approach because the optical depth of soft X-ray becomes $\sim 1$ when the density reaches $\nh \sim 10^4$~\hcc\ in our haloes, before the formation of a disc. We expect X-ray FB in high-density gas to be negligible (because self-shielded) in any direction and thus our results are insensitive to this choice. In the SF simulations, the column density is
\begin{equation}
    N_{c} (r) = N_{box} (r) + N_{out}.
\end{equation}
Because the virial radius is larger than the size of the box (2~pc), we assume that the column density at $r > 1$~pc ($N_{out}$) is constant in time since $\nh = 10^7$~\hcc. We instead update the column density inside the box ($r \leq 1$~pc, $N_{box}$) at every coarse time-step ($\sim 13$~yr). 

We also include shielding of the LW background. The LW shielding includes self-shielding of \hm\ and HD and shielding of HD by \hm. For the self-shielding factor of \hm, we use equation~(7) in \citet{wolcottgreen2019}. In our work, this factor is denoted by $f_{WGH} (\nchm,T)$ where $\nchm = \nhm \Delta x$ is the \hm\ column density and $T$ is the temperature of the cell. We provide a discussion on the column density in Section~\ref{sec:transfer}. We use the same formula for the HD self-shielding but with the column density replaced by that of HD. The shielding of HD by \hm\ is computed with equation~(12) in \citet{wolcottgreen2011a}. The total shielding factor of HD is the multiplication of the two factors. We did not implement the shielding of \hm\ by HD because the HD density is lower by several orders of magnitude. We study the importance of X-ray self-shielding on the long-term evolution ($\sim 30$~kyr) of the protostars forming in Halo~2, in Appendix~\ref{app:shd}. Note that \citet{hummel2015} evolved the simulations for only $\sim 5$~kyrs, and our test suite in Paper~II also covered a shorter time. However, the importance of self-shielding may vary depending on the growth history of the chosen halo as suggested in \citet{hummel2015}.
\begin{table}
    \caption{Energy bins.}
    \begin{threeparttable}
    	\centering
    	\label{tab:bin}
    	\begin{tabular}{ | c | c | c | l | }
		    \hline
            bin & $h_{P}\nu_1$~(eV) & $h_{P}\nu_2$~(eV) & Roles \\
		    \hline
            1 & 11.20 & 13.60 & \hm\ dissociation \\
            2 & 13.60 & 15.20 & \HI ionization \\
            3 & 15.20 & 24.60 & \HI, \hm\ ionization \\
            4 & 24.60 & 54.42 & \HI, \hm, \HeI ionization \\
            5 & 54.42 & $\infty$ & \HI, \hm, \HeI, \HeII ionization \\
            \hline
	    \end{tabular}
    \end{threeparttable}
\end{table}

\subsection{Sink Formation and Accretion}
\label{sec:sink}
In hydrodynamics simulations the Jeans length must be resolved with a minimum number of cells to prevent artificial fragmentation \citep{truelove1997} and better resolve turbulent motions. However, as the Jeans length approaches the minimum cell size ($\Delta x$) in the simulation, this criterion fails. To circumvent this issue in Paper~I and Paper~II we artificially reduced the cooling rate of Jeans unstable cells near the resolution limit, as also done in previous works \citep{hosokawa2016,hirano2017}. Because of the nearly adiabatic collapse heating, the gas stops collapsing forming clumps that we regard as Pop~III protostars. This method is not ideal to track the evolution of individual clumps and implement RFB from them. For this reason in this work we utilise sink particles, that we refer to as stars or protostars. We follow the recipe in the public version of \ramses\ \citep{bleuler2014} with minor modifications. The built-in clump finder \citep{bleuler2015} identifies gas clumps and we check if each clump satisfies the following,
\begin{enumerate}
    \item The clump peak density is $n_H>n_{sink} = 10^{12}$~\cc.
    
    \item There is no pre-existing sink within $2 r_{sink} = 2 \times N_{Sink} \Delta x$.
    
    \item The flow is converging, $\nabla \cdot \overrightarrow{v}<0$.
\end{enumerate}
If all conditions are met for a given clump then we create a sink particle at its density peak. In Appendix~\ref{app:clump} we compare the star formation with clumps and sink particles.
\begin{figure}
    \centering
	\includegraphics[width=0.48\textwidth]{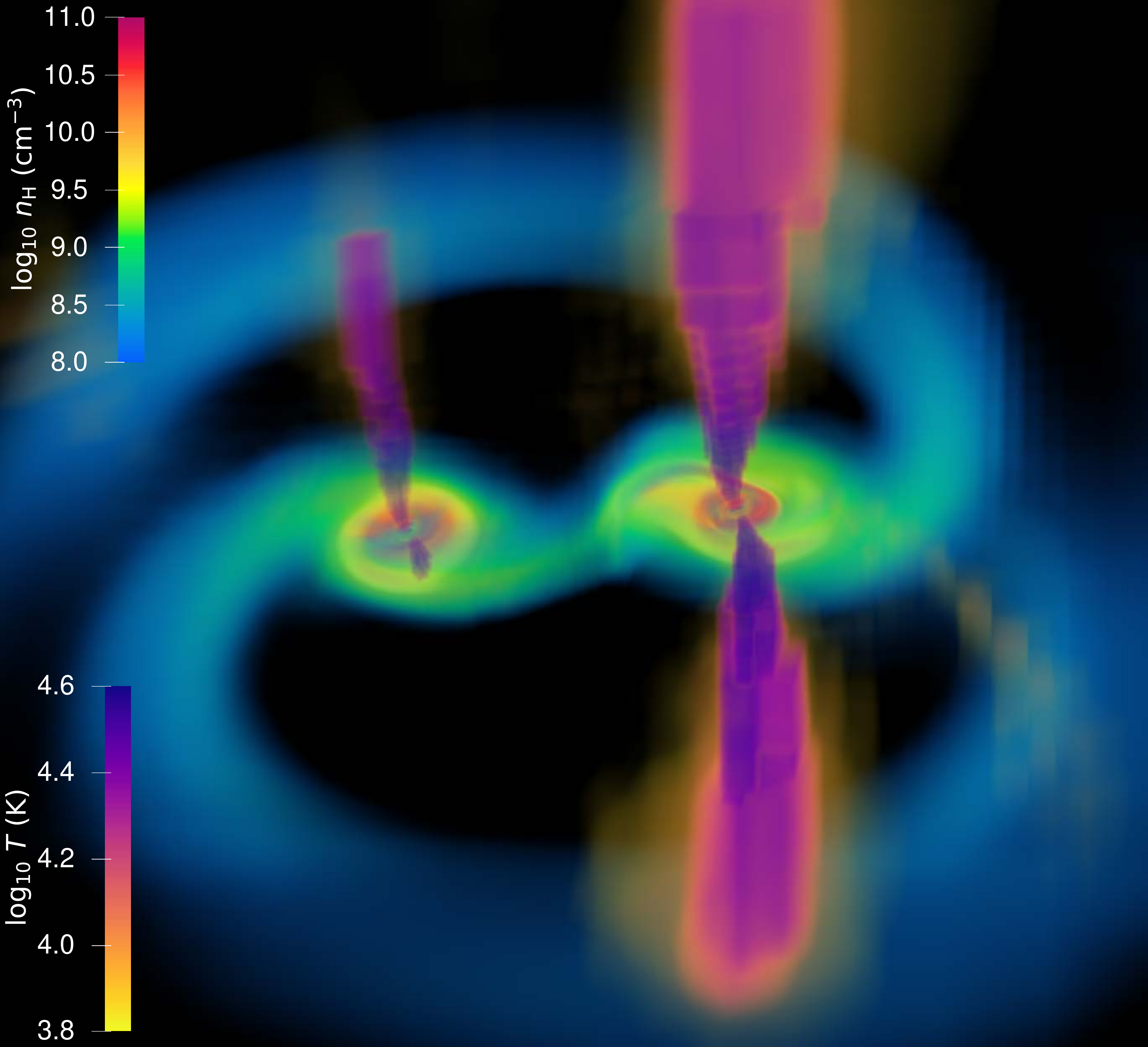}
    \caption{Rendered image of a nearly equal mass binary Pop~III protostar and disc in Run~F. Two stars accrete gas from their own circumstellar mini-discs, and the protostellar UV radiation produces two hot bipolar \HII regions and outflows with velocities $\sim 100$~km/s. The mini-discs orbiting the common CoM sweep and accrete the gas in the main disc. The field of view is $\sim 10000~{\rm AU}~\times~10000~{\rm  AU}$.}
    \label{fig:3d}
\end{figure}

In the literature, the sink radius is chosen to be a multiple of the smallest cell size ($N_{Sink} \Delta x$), with values of $N_S$ ranging from 4 to 16. We choose $N_{Sink} = 8$ and thus $r_{sink} = 101$~AU. If a gas cell is within the radius of a sink particle and its number density exceeds $n_{sink}$ the excess mass ($(n - n_{sink}) \Delta x^3$) is dumped into the sink. Two sink particles merge if they are closer than $2r_{sink}=202$~AU. Cells within $1.5 r_{sink}$ from the centers of any sink particle are fully refined to guarantee accurate calculation of hydrodynamics and radiative transfer near sinks. In Appendix~\ref{app:sink} we validate our choice of $n_{sink}$ and provide test results with different $N_{Sink}$.

\subsection{Photon Injection}
\label{sec:inj}
We utilise the model of protostars calculated by \citet{hosokawa2009} and \citet*{hosokawa2010} to obtain the amount of photons emitted by sink particles. These authors traced the evolution of accreting protostars with a 1-D stellar evolution code and tabulated the radius and luminosity (and thus effective temperature, $T_{eff}$) as a function of stellar mass and gas accretion rate. In our simulations, the luminosity of a sink particle in each frequency bin is,
\begin{equation}
    Q = 4 \pi R^2 \int_{\nu_1}^{\nu_2} \frac{\pi B_\nu(T_{eff})}{h_{P} \nu} {\rm d} \nu,
\end{equation}
where $R$ is the effective radius of the model star and $T_{eff}$ is its effective temperature, assuming it has a Plank spectrum $B_\nu(T_{eff})$. At each time-step of the simulation, the values of $R$ and $T_{eff}$ are interpolated from the table as a function of the mass and accretion rate of the sink particle. The lower and upper bounds of frequency ($\nu_1$ and $\nu_2$) in each bin are given in Table~\ref{tab:bin}. We do not include luminosity due to gas accretion. In a time-step $\Delta t$, we inject $Q \Delta t$ photons around each sink particle. Throughout the paper, we use the term LW and FUV (far-ultraviolet) interchangeably for the radiation in the first energy bin. We call the radiation in the other bins either ionizing or EUV (extreme-ultraviolet) radiation.

For a Pop~III star to form the core density should exceed $10^{20}$~\cc\ \citep{omukai2001}. This density cannot be resolved in our simulations so we need a sub-sink recipe to model photon injection from the sinks. Instead of assuming that all cells within the sink are sources of photons, we scatter photons equally on the sink surface, with fluxes in the outgoing radial direction. The specific intensity of photons launched from the surface in each direction is further reduced by the escape fraction $f_{esc}$:
\begin{equation}
f_{esc}=
\left\{
	\begin{array}{ll}
		0 & {\rm if}~n > n_{inj} \\
		f_{WGH} (\nhm r_{sink},T) & {\rm if}~n \leq n_{inj} ~{\rm (FUV)} \\
		1 & {\rm if}~n \leq n_{inj} ~{\rm (EUV)} \\
	\end{array}
\right.,
\label{eq:esc}
\end{equation}
where $n_{inj} = 0.1 n_{sink}$ is the photon injection density threshold. We assume $1-f_{esc}$ photons are absorbed within the sink in the path from the central star to the sink surface. With this recipe, the radiation propagating in directions close to the plane of the disc is efficiently suppressed, while a significant fraction of EUV and LW photons escape in the vertical direction. Cells with injected photons have photon flux $n_{ph}c_{red}\hat{r}$ where $c_{red}$ is the reduced speed of light and $\hat{r}$ is the unit displacement vector from the centre of the sink. Tests leading to the adoption of this recipe and the density threshold for photon injection ($0.1 n_{sink}$) are shown in Appendix~\ref{app:inj}. The results of the simulations can be sensitive to the model of photon injection, especially with respect to the naive assumption that all photons are injected in the centre or are spread within the sink. 

\begin{figure*}
    \centering
	\includegraphics[width=0.95\textwidth]{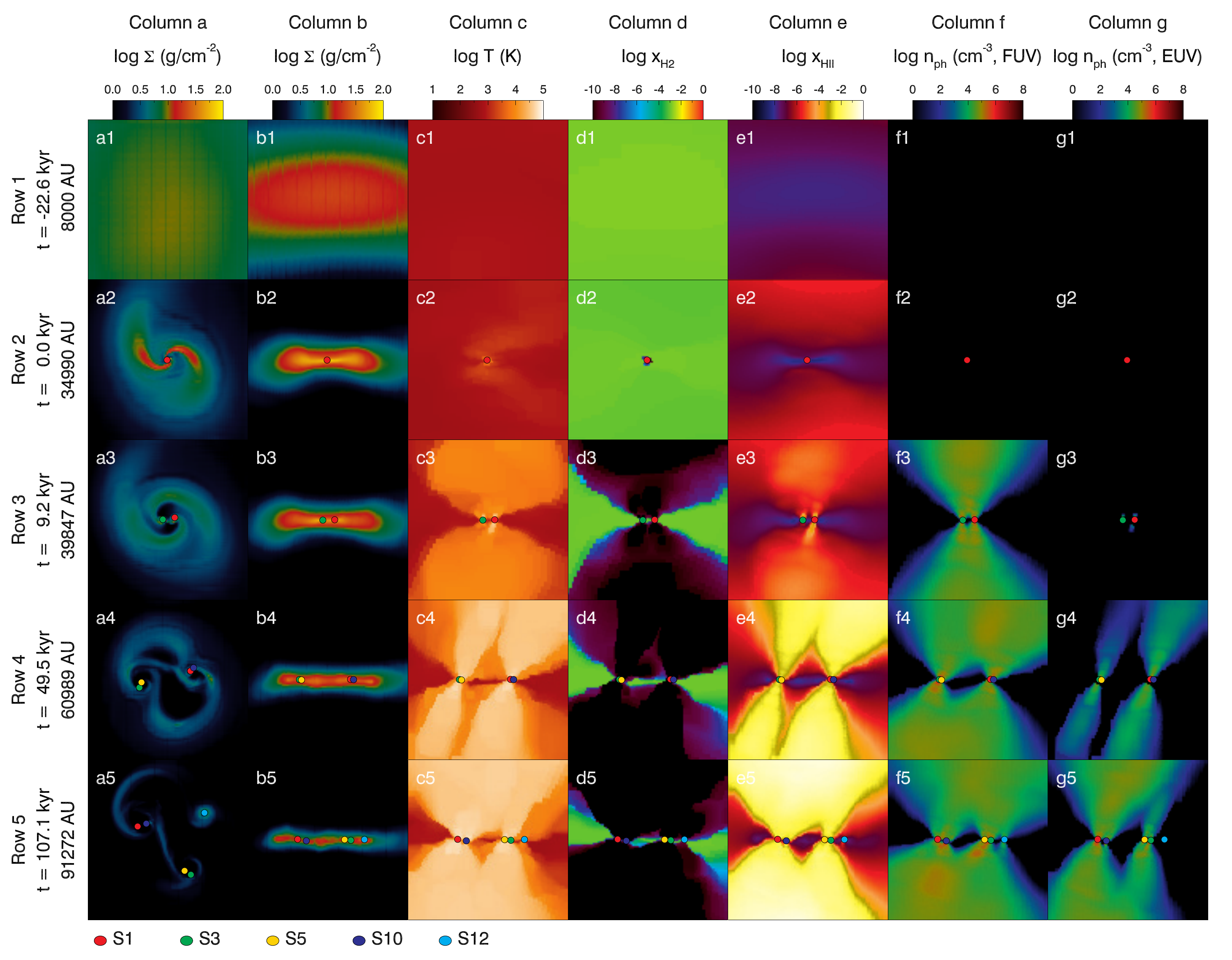}
    \caption{Snapshots of Run~F at different times (Row 1-5). The scale of each frame, shown on the left, increases as the disc and the protostars' orbits expand. We define $t=0$ when the first sink particle forms. We also plot protostars as filled circles (see the bottom of the figure for the labels). Column~a shows the face-on view while the other columns (Column~b to g) show the edge-on view. The label on each panel refers to the column and the row. Column~a: Gas surface density. Through the initial fragmentation, the system starts with a binary (a3). The circumstellar discs fragment and a hierarchical binary forms (a4). Column~b: Gas column density. The gas flattens (see b1 and b2) and a disc forms. All protostars are in the disc plane. Column~c: Slice through the disc center showing the gas temperature. Note that the heated regions are traced by the radiation field. LW radiation heats the gas up to $T \sim 10^3$~K and the heating by ionizing radiation (EUV) results in $T \gtrsim 5\times 10^4$~K. Column~d: Slice of \hm\ fraction. LW photons dissociate \hm. The opening angle of the photodissociation region increases with time. Column~e: Density weighted \HII fraction. The radiation from protostars creates bi-conical \HII regions. Column~f~\&~g: Projected LW and ionizing UV photon density. Two massive sink particles emit a copious amount of photons. The image shows the precession of mini-discs. A movie version of this figure is available in the supplementary material.}
    \label{fig:feedback}
\end{figure*}

\subsection{Radiative Transfer}
\label{sec:transfer}
Radiation transfer in \ramsesrt\ is implemented using the M1 closure method \citep{rosdahl2013}, with photon number density and flux in each cell computed locally. For hydrogen and helium ionizing radiation, we follow the prescriptions in the publicly available version. Unlike the radiative transfer of ionizing UV/X-ray radiation in which the opacities are broad in frequency, opacity in the LW band is produced by absorption by numerous narrow lines with widths and centres that depend on temperature and kinematic of the gas  \citep[e.g., see][]{RicottiGS:2001}, thus its calculation is computationally expensive. A widely used approximation is to assume that the gas is optically thin to LW radiation and correct the \hm\ dissociation rate with the self-shielding factor \citep{draine1996}. For the self-shielding factor we use the up-to-date equations in \cite{wolcottgreen2019},
\begin{equation}
    f_{shd} = f_{WGH} (\nchm, T).
    \label{eq:sh}
\end{equation}
This factor, however, is a function of column density $\nchm$ and its calculation is also cost-prohibitive in a hydrodynamics simulation. For this reason, the column density is often replaced by the local column density computed using the Jeans length, velocity gradient, or density gradient \citep*{wolcottgreen2011b} although their usage still requires cautions \citep{chiaki2022}. We follow a mixed local/global approach to calculate the opacity in the LW bands. We calculate the self-shielding locally for the gas inside each cell using equation~(\ref{eq:sh}) with $\nchm = \nhm \Delta x$, as in \citet{katz2017}, but we also account for global opacity in the LW frequency bin in the M1 closure method, assuming an effective cross-section $\sigma_{LW} = 2.47 \times 10^{-18}$~cm$^2$ for H$_2$ gas \citep{katz2017}. We also include LW photon absorption by atomic hydrogen with $\sigma_{Lyman} = 5.23 \times 10^{-25}$~cm$^2$  \citep{wolcottgreen2011a,jaura2022}. The effect of the shielding by \HI is discussed in Appendix~\ref{app:inj}. 

To speed up the calculations, we use the reduced speed of light approximation. One must use this approximation with care in cosmological simulation \citep{gnedin2016} but usage in a high-density regime like in SF simulation is valid. The reduced speed of light is $c_{red} = 10^{-3}c = 300~\kms$ in this work. We validate this choice in Appendix~\ref{app:rsla}. The computation speed does not increase further with a lower value, as the maximum outflow speed is $\sim 100~\kms$. GLF Riemann solver is recommended for cosmological simulation \citep{rosdahl2013} but we use HLL Riemann solver \citep*{harten1983} motivated by the shadowing effect. However, note that both solvers show a similar degree of diffusivity when radiation is transferred in a diagonal direction and therefore our results are independent of this choice. Finally, we provide test results with FUV- and EUV-only  simulations in Appendix~\ref{app:UV} to explore the effect of each radiation. We also test the resolution effect in Appendix~\ref{app:res}.

\begin{figure}
    \centering
	\includegraphics[width=0.48\textwidth]{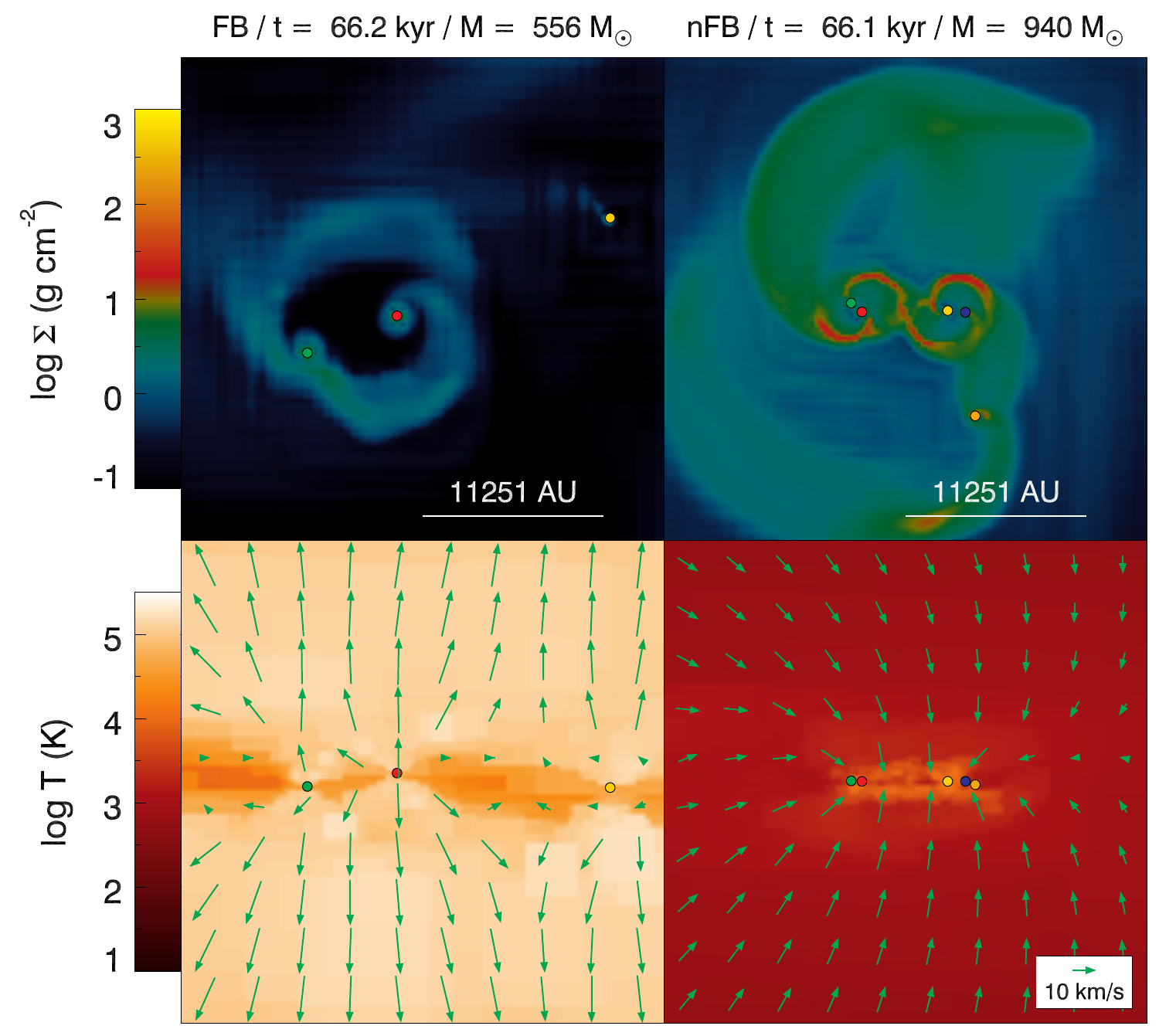}
    \caption{Face-on gas surface density (top) and edge-on temperature (projected, bottom) in Run~A. We present FB (left) and noFB (right) for comparison. The time and total mass are shown above each density plot. The time is nearly at the end of the FB run. The field-of-view is $30000~{\rm AU} \times 30000~{\rm AU}$. On top of the temperature plots we plot the velocity field (green arrows). The size of an arrow scales with the speed logarithmically and the maximum speed is $\sim 100$~km~s$^{-1}$. Protostars are shown by the filled circle.}
    \label{fig:snapa}
\end{figure}


\section{Results}
\label{sec:results}
\subsection{Formation of Protostars in Primordial Discs}
\label{sec:mass}
It has now been shown by several authors that a primordial gas discs fragment to form multiple Pop~III protostars \citep{machida2008,stacy2010,clark2011}, but early studies evolved the protostars for short times with respect to the accretion timescale determined by RFB. Recently \citet{sugimura2020} was able to simulate a multiple stellar system until protostellar RFB shuts down accretion on the protostars ($\sim 120$~kyr). In our simulations we also find that the formation of multiple Pop~III stars is ubiquitous, and the growth of protostars is quenched by RFB. Fig.~\ref{fig:3d} shows the 3D rendering of one of our simulations, in which a binary star emitting radiation produce bipolar outflows. Protostars accrete gas from their own circumstellar mini-discs, emitting FUV and EUV photons. These photons create hot bipolar \HII regions in the vertical direction, that can be tilted with respect to the main disc and asymmetric as observed for the left star, where the south outflow is weaker than the north one. While the stars orbit each other, their circumstellar discs accrete  gas from the main disc and create gaps and extended arms as also seen in idealised simulations of circumbinary disc (e.g., \citealp*{tang2017}; \citealp*{moody2019}; \citealp{dittmann2022}).
In this section we show the results of simulations of disc fragmentation and Pop~III star formation for various radiation backgrounds in two representative dark matter minihaloes.
Fig.~\ref{fig:feedback} shows the central gas evolution for Run~F at different epochs (a movie of the simulation is available in the online version of the paper). The qualitative evolution is similar across different simulations, even though the masses and multiplicity of the stars show trends with the intensity of the X-ray background. 

\textbf{\emph{Typical formation scenario.}} The gas cloud contracts and flattens to form a disc and protostars (in Fig.~\ref{fig:feedback}, panels~a1, a2 show the face-on view and panels b1, b2 the edge-on view). Soon after the first protostar appears ($t=0$), three stars form near the disc centre at $t \sim 1$~kyr through the initial fragmentation of a prominent barred spiral arm structure developing in the relatively massive disc around the low-mass central protostar \citep{kimura2021}. Two of them merge with each other and only two nearly equal mass stars (S01) and (S03) are left orbiting the centre-of-mass (CoM) of the system (see panels~a3/b3). Each of these protostars has circumstellar mini-discs which accrete mass from the larger disc. In all the simulations these nearly equal mass binaries (with stars of about 100~M$_\odot$), migrate outward from a few 100~AU separations at formation, to 10,000~AU by the end of the simulation at $t\sim100$~kyr.  At time $t \sim 15$~kyr, the mini-disc around S03 becomes gravitationally unstable through mass accretion and fragments \citep{sugimura2020} to form S05. Similarly, at $t \sim 24$~kyr, the fragmentation of the circumstellar disc around S01 forms S10. At this time the system is a hierarchical binary where two binary stars, with nearly equal total mass, orbit each other (panel~a4/b4). The protostars in each binary system have their own circumstellar mini-discs. As discussed in this section and the upcoming paper (in prep) these discs play an important role in the growth and migration of protostars. The stars are getting further from their common CoM with time and the binaries also are getting further as the disc expands. At later times (panel a5/b5) S12 forms at Lagrange points L4 (or L5) of the main hierarchical binary. Afterward, star formation becomes more stochastic due to the large multiplicity, and zero-metallicity low-mass stars can form when rapidly ejected from the disc. 

Fig.~\ref{fig:feedback} also shows that the onset of RFB changes the environment around the protostars dramatically. As the protostars grow in mass, they become more luminous and their spectra become harder. In the beginning, LW photons from the stars propagate through the gas and photodissociate \hm\ in the vertical direction as shown in panels~(d3), showing the \hm\ fraction and (f3), showing the number density of LW photons. For the first $10-20$~kyr the ionizing radiation remains trapped by the dense neutral gas in the disc (panel~g3) and the \HII region is trapped (panel~e3). Later, ionizing radiation breaks through the gas in the disc forming bi-conical \HII regions (see panels e4/e5 and g4/g5). The radiation field is dominated by the two protostars that form earlier because the ones forming later from the fragmentation of the mini-discs are less massive. 
Ionization and photodissociation by radiation heat the gas: \hm\ dissociation due to LW photons heat the gas above the disc in a broad region to $T \sim 1,000$~K, creating an outflow. Note that the opening angle of the bi-conical regions where the gas is heated to 1,000~K  (see panels~c3 to c5), is similar to the opening angle of the LW radiation field (panel~f3 to f5). As ionizing photons escape the disc they form bi-conical \HII regions with a narrower opening angle than the LW counterpart (Column~g). Photoionization heating increases the temperature to $T \gtrsim 5 \times 10^4$~K in these narrower conical regions (Column~c). The EUV heating enhances the bi-polar outflow velocity as seen in Fig.~\ref{fig:snapa}, thereby suppressing the growth of the stars.
\begin{figure}
    \centering
	\includegraphics[width=0.48\textwidth]{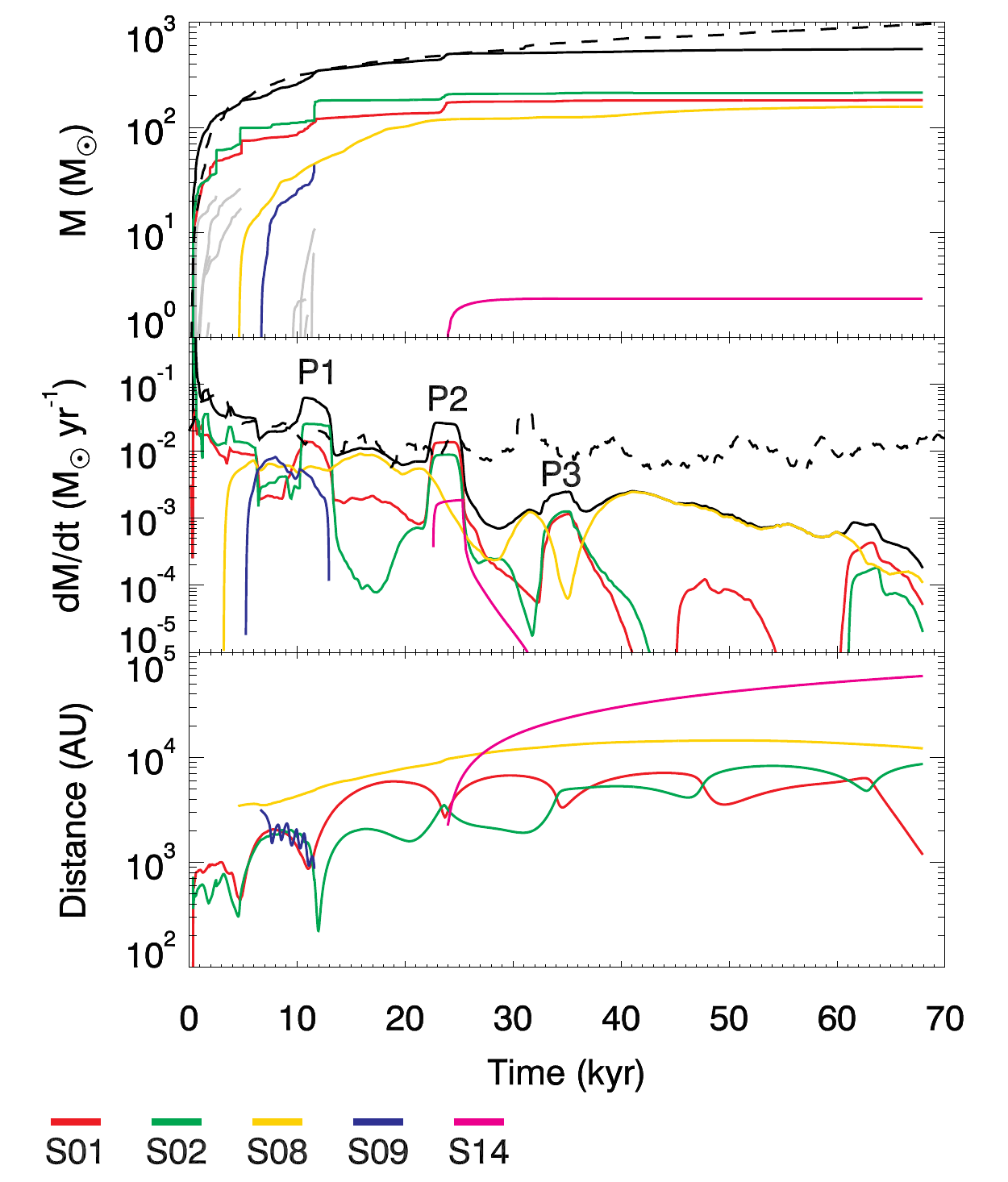}
    \caption{Top: Mass in protostars with time for Halo2 with $J_{X0,21}=0$ (Run~A). Solid lines are in the fiducial case and the dashed line is in noFB case. The total mass is shown in black and individual stars are shown in different colors and grey with long-lived ones highlighted (see bottom). Note that the mass of low mass S14 (magenta) remains constant without merging with other stars. Middle: Accretion rates in $\msyr$. Similar to the top panel but we plot only the long-lived ones. We mark accretion rate peaks with P1, P2 and P3. Bottom: Distance from CoM. The final distance of S14 is $\sim 6 \times 10^4$~AU.}
    \label{fig:massa}
\end{figure}

\subsection{Roles of FUV and EUV radiation feedback} 
\label{sec:sima}

In paper~I and II we found that the total mass in Pop~III stars depends on the accretion rate as
\begin{equation}
M_{pop3,tot} \propto (dM/dt)^{0.7},\label{eq:tsf}
\end{equation}
in agreement with HR15. If we define the star formation quenching timescale $\tau_{SF} \equiv M_{pop3,tot}/(dM/dt)$, we expect $\tau_{SF} \propto M_{pop3,tot}^{-0.3}$: i.e.,  higher mass protostars have a stronger feedback effect than lower mass stars. 

Without X-ray irradiation the accretion rate is highest (because X-rays, cool the primordial cloud core by enhancing H$_2$ and HD formation), thus stars grow rapidly and are more massive than in cases with X-ray irradiation. Hence, we expect a faster suppression of the mass growth of Pop~III stars due to RFB in runs without X-rays. To test this model, we explore the evolution of protostars including and excluding RFB effects in Halo~2, when irradiated by different intensities of the radiation background.\footnote{In Paper~I we found that Halo~2, because of its regular accretion history, shows the clearest trend with varying the intensity of the X-ray background.}
In Appendix~A7 we test the relative importance of EUV and FUV feedback by running the same simulation with both EUV + FUV feedback, only EUV, only FUV, and noFB. We find that FUV and EUV feedback have comparable effects on the suppression of the accretion rate onto Pop~III stars.

Fig.~\ref{fig:snapa} shows the gas surface density (face-on) and temperature (edge-on) of the FB (left) and noFB (right) Run~A ($J_{X,21}=0$) at $t \sim 66$~kyr. The overall gas density is lower in the FB run because RFB suppresses the gas supply and available gas is already consumed by the existing stars. The protostars in this simulation are more massive ($M \sim 100-200~\msun$) than those in Run~D ($M < 100~\msun$) seen in Fig.~\ref{fig:feedback}, hence the bipolar \HII regions reach temperatures of nearly $10^5$~K. As seen in the temperature plots an outflow with velocity $\sim 100~\kms$ in the vertical direction develops and reverses the gas motion (arrows in the bottom left panel), that would otherwise be accreting onto the disc and stars (bottom right). Generally, we find across all simulations that FB reduces the total mass in stars both at intermediate times and even more at late times, since the total mass in stars continues to increase with time at a nearly constant rate in the noFB cases. The multiplicities in the last snapshot ($N_{final}$) are 4 and 7 showing fewer stars in the FB run. Again, at an earlier time ($t \sim 68$~kyr), the number of stars is 4 and 5 in both runs, hence the difference is smaller.
\begin{figure*}
    \centering
	\includegraphics[width=0.48\textwidth]{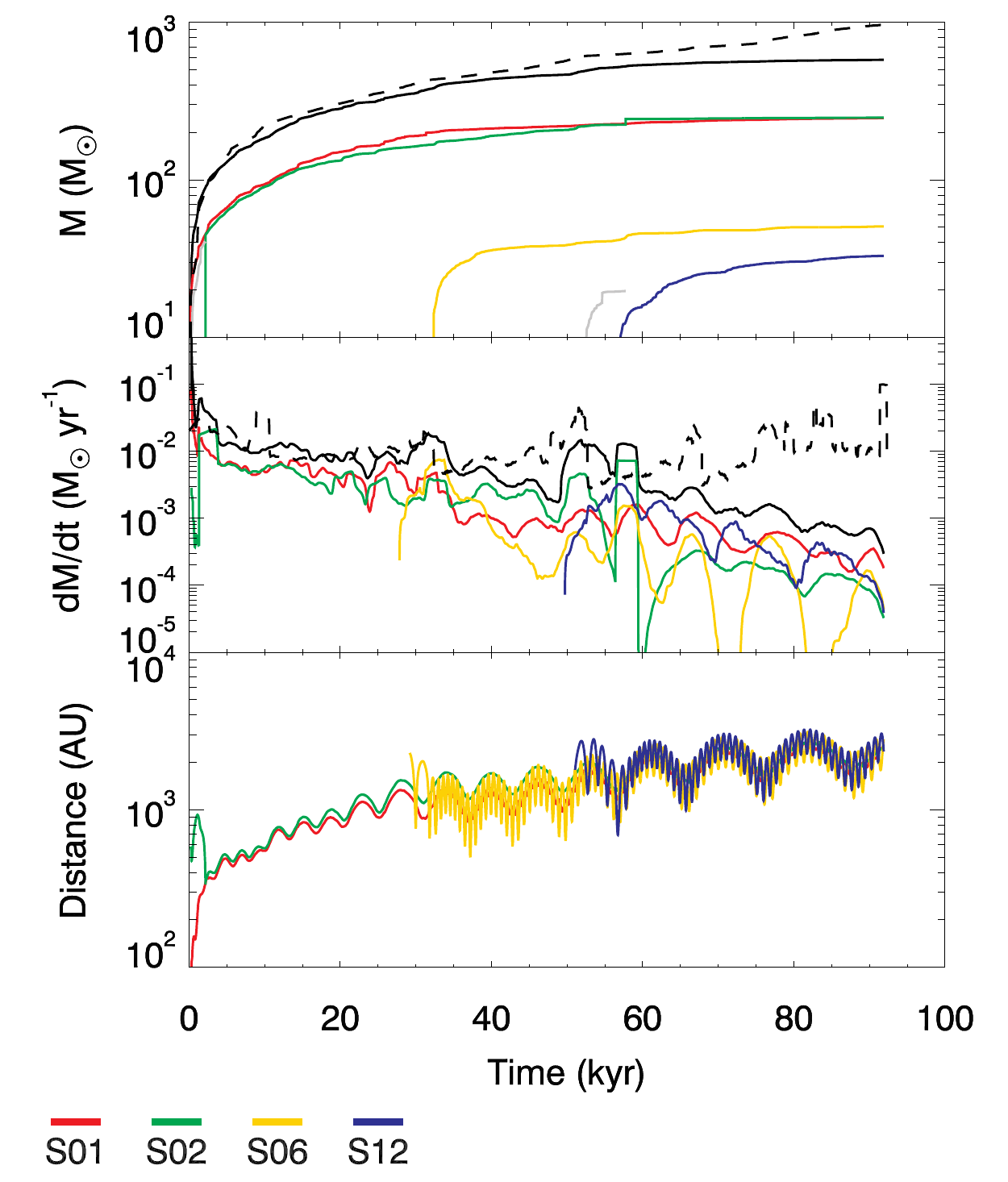}
    \includegraphics[width=0.48\textwidth]{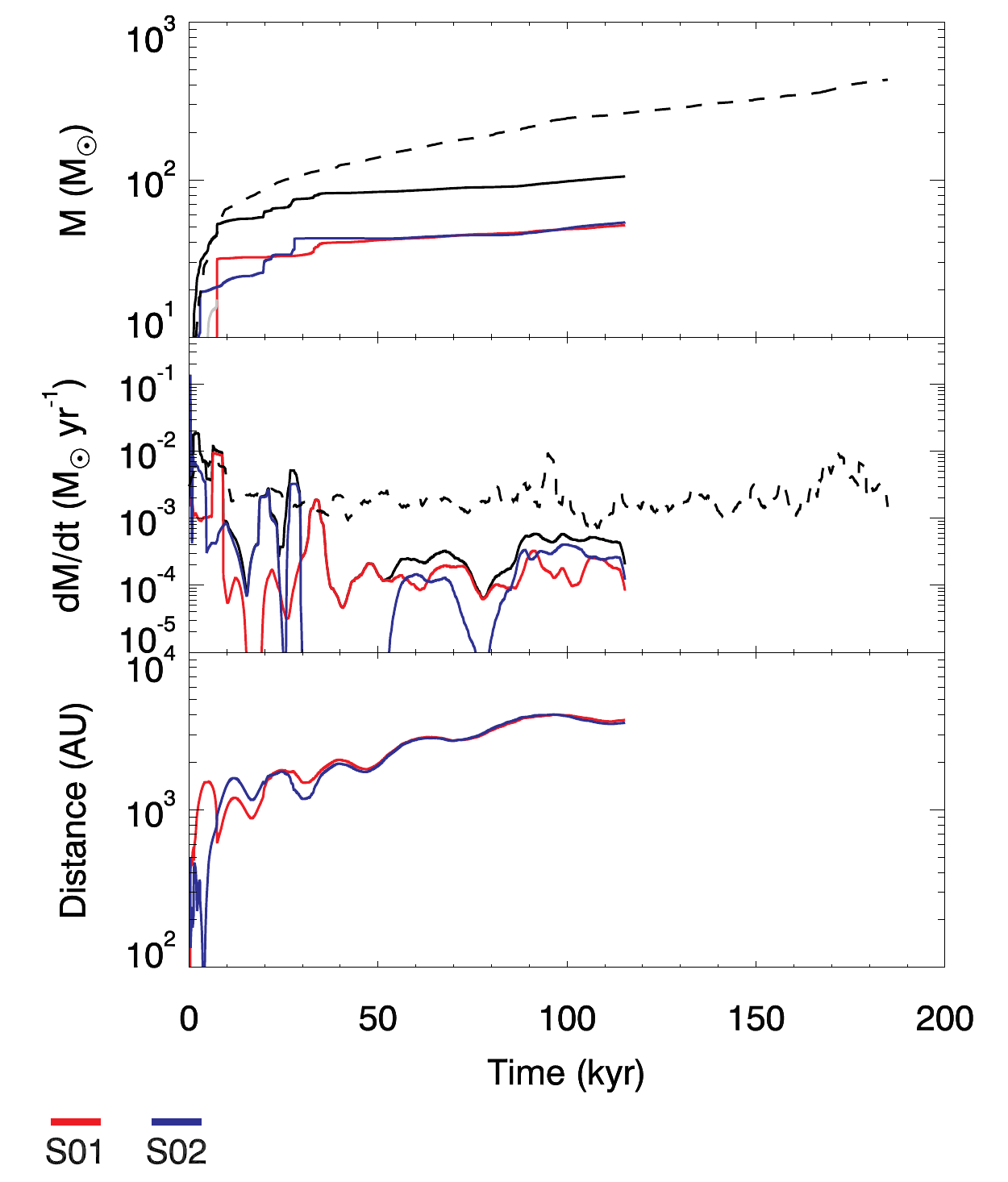}
    \caption{Left: Same as Fig.~\ref{fig:massa} but for $J_{X0,21}=10^{-4}$ (Run~C). Note that the middle and bottom panels shows only long-lived stars. Right: Same as the left panel but for $J_{X0,21}=10^{-2}$ (Run~E). Four protostars form in total but only two of them do not merge with each other for $\gtrsim 100$~kyr.}
    \label{fig:massc}
\end{figure*}

\subsection{Modulation of the Accretion Rate and Feedback due to Eccentricity of the Orbits}
\label{sec:halo2}

In Fig.~\ref{fig:massa} we plot the masses, accretion rates and distances from the CoM as a function of the time of the stars forming in Halo~2 for the case without X-ray and LW backgrounds (Run~A). Single stars are shown by colored lines for the fiducial simulation, while the total mass for the noFB simulation (black dashed lines) is compared to the fiducial case (black solid lines). The bottom panel shows that the two most massive stars are in a binary with rather eccentric orbit ($e \approx 0.8$) as shown by the bottom panel. This is somewhat surprising given that the stars are still deeply embedded in a gas-rich quasi-Keplerian disc. However, the disc is highly inhomogeneous due to the presence of prominent arms and bar features. We will see in a follow-up paper that these non-axisymmetric features are responsible for the outward migration of the stars and allow initially eccentric orbits to avoid circularizing.

\noindent
\textbf{\emph{Eccentricity-induced periodic growth.}}
RFB suppresses the growth of protostars as suggested in previous studies \citep{hosokawa2011}. The accretion rate in the noFB run is almost constant for at least 100-200~kyrs, while the accretion rate slows down in the run with RFB (fiducial run). However, in the fiducial run there are times when ${\rm d}M/{\rm d}t$ suddenly increases and sometimes the accretion rate exceeds that in the noFB case. We mark these episodes with the labels P1, P2 and P3 in the middle panel of Fig.~\ref{fig:massa}. At each peak, the S01-S02 pair (red and green) grows rapidly and this occurs at the minimum separation of the two stars (i.e., at pericenter distance). As shown in the bottom panel, the binary has an eccentric orbit, and the distance between S01 and S02 has a minimum at P1, P2 and P3. The periodic increase of the accretion rate is likely produced by the tidal interaction of the circumstellar mini-discs, funnelling the gas to the stars and thus boosting the growth of the protostars. The accretion rates at these peaks decrease with time due to the gas depletion and the third peak is lower than the noFB case. Unlike the case of AGN feedback in galaxy merger simulations \citep*{park2017}, tidal interaction does not always lead to stronger feedback. At P1 and P2 the tidal interaction reduces the ionizing luminosities of S01 and S02 by several orders of magnitude, weakening RFB. This is because protostars are modeled to enter a supergiant phase (and therefore their effective temperature drops) when they have high accretion rates: ${\rm d}M/{\rm d}t \gtrsim 10^{-2}~\msyr$ \citep{hosokawa2009,hosokawa2010,hosokawa2016}. As seen in the figure, the accretion rates of these two stars are similar to this critical value at P1 and P2. At P3, however, the stars do not enter the giant phase and their luminosities increase for a short period of time. While RFB from stars S01 and S02 are shut down during the close encounters, the ejected star S08, keeps radiating and produces an outflow.

In Paper~I and Paper~II we used the empirical relationship from \citet{hirano2014} and HR15 to determine the final masses of Pop~III stars (RFB was not included in those simulations). In this work we simply define the mass in stars as the one in the final snapshot of the simulation, denoted $M_{final}$. The total mass in stars in Run~A is $557~\msun$ as seen in Table~\ref{tab:sim} and is clearly lower than the noFB case at the same time ($1086~\msun$). However, even at the end of our simulations, RFB has not fully evaporated the gas disc, especially the gas in the circumstellar mini-discs, and therefore the stars keep growing at a very slow rate. For this reason, $M_{final}$ provides only a lower limit to the mass of the stars, even though we do not expect that their masses would increase significantly even if we evolved Run~A further. As seen in Fig.~\ref{fig:feedback}, little dense gas is left due to the strong RFB and the total accretion rate in the last snapshot is $\sim 10^{-1}~\mskyr = 10^{-4}~\msyr$ (Fig.~\ref{fig:massa}): if stars accrete the gas at the same rate for the next 100~kyr, the mass increase would only be $10~\msun$ and the total mass growth less than 10 percent. On the other hand, in the noFB simulations, the accretion rate would remain constant  at about 100 times higher rate, and the total mass would roughly double reaching 2000~M$_\odot$.

\noindent
\textbf{\emph{Hierarchical binaries and outward migration.}} 
The bottom panel of Fig.~\ref{fig:massa} shows the distances from the CoM of long-lived stars. Eight protostars form initially in the inner disc, but they merge with each other producing a triple system of stars: S01, S02 and S08. Star S08 (in yellow) is ejected via gravitational interaction and only S01 and S02 orbit each other in the end. The orbits of these stars initially migrate outward, but at later times they reach a constant radius. The orbit of the S01-S02 binary is highly eccentric with the pericenter distance $\sim 1000$~AU and the apocenter distance $\sim 8000$~AU. Although new protostars form while S01 and S02 orbit each other, they all merge with S01 and S02. At each pericenter, tidal interaction between the two mini-discs increases and tidal tails and a gas bridge/bar becomes more prominent, somewhat reminiscent of features observed in galaxy merger simulations \citep{toomre1972,cox2008}.
\begin{figure*}
    \centering
    \includegraphics[width=0.48\textwidth]{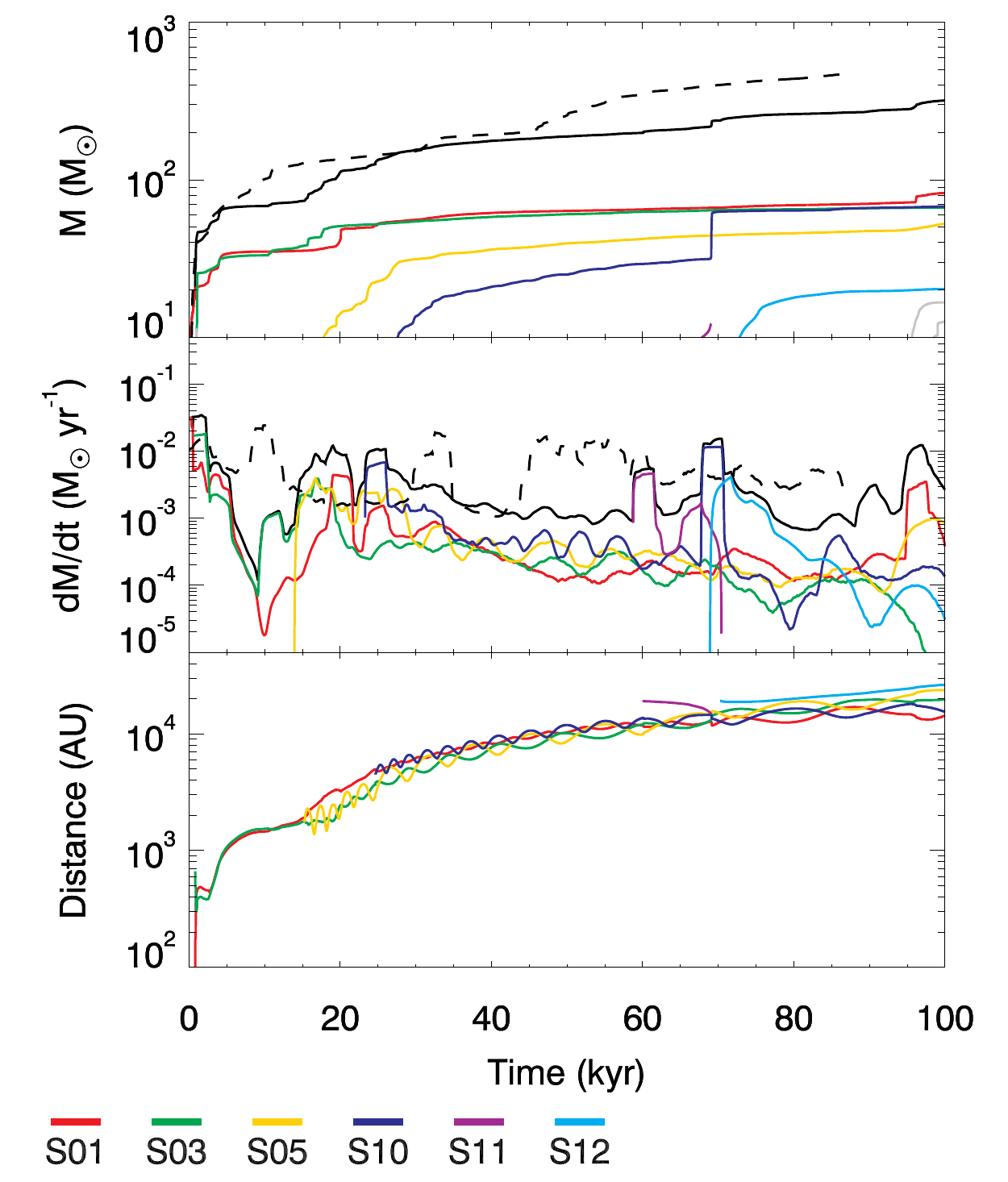}
	\includegraphics[width=0.48\textwidth]{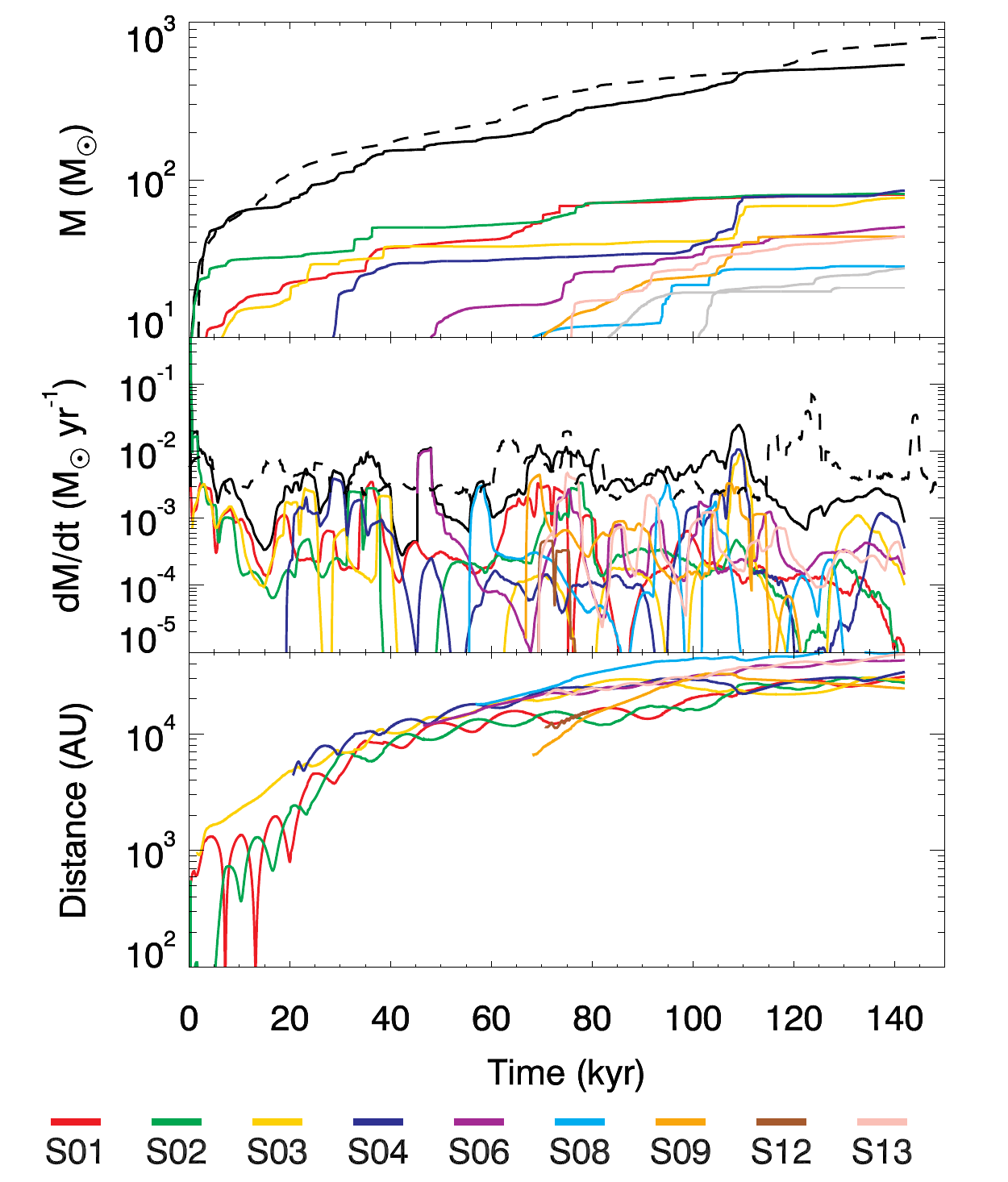}
    \caption{Left: Same as Fig.~\ref{fig:massa} but for Halo3 with $J_{X0,21}=0$ (Run~F). Right: Same as the left panel but for  Same as Fig.~\ref{fig:massa} but for $J_{X0,21}=10^{-5}$ (Run~G).}
    \label{fig:massg}
\end{figure*}
At $t \sim 24$~kyr, S14 (magenta in Fig.~\ref{fig:massa}) forms out of a tidal tail and is ejected from the system (see the bottom panel). At the time of its formation, the gas is significantly depleted, so it does not have enough gas to accrete and thus remains with a relatively small mass ($M\approx 2~\msun$). Although its mass is greater than $1~\msun$ this supports the possibility that low-mass Pop~III stars can be observed today \citep{clark2011, greif2011, stacy2013}. While previous works simulated the protostellar disc for a relatively short time (a few kyr), our simulation follows the orbit of unbound low-mass protostars for about 100~kyr, as in \citet*{susa2014}. Two protostars are also ejected in the noFB case, but they are more massive ($10$~and $4~\msun$).

\subsection{Pop~III masses and multiplicity in different X-ray backgrounds}

\textbf{\emph{Multiplicity.}} We confirm the results in Paper~II that in an X-ray background the disc is more gravitationally stable and the multiplicity of stars is reduced. For example, for $J_{X0,21}=10^{-4}$ (Run~C),  the mini-discs do not fragment for $\sim 29$~kyr since the initial formation of a central binary as seen in the top panel of Fig.~\ref{fig:massc}~(left). At $t \sim 29$ and $52$~kyr, S06 and S12 form in the circumstellar mini-discs of S01 and S02, respectively, forming a hierarchical quadruple system. The final multiplicity is equal to that found for $J_{X0,21}=0$, mostly because in the latter several sinks merge with each other (see Table~\ref{tab:sim}). For stronger X-ray irradiation, $J_{X0,21}=10^{-2}$, the second fragmentation occurs similarly at $t \sim 25$~kyr but the new sink quickly merges with the existing star. No further fragmentation occurs for the next $\sim 90$~kyr (see the top right panel of Fig.~\ref{fig:massc}) and the system ends up being a binary. Other simulations with non-zero X-ray backgrounds (Run~B and Run~D) have multiplicity 3 and 2, respectively. Although Run~C seems to deviate from the general trend (we attribute it partially to the larger halo/disc mass produced by the delayed collapse in a non-zero LW background), the other runs agree with the expectation from Paper~II: X-rays make the disc more Toomre-stable, reducing the multiplicity.

\noindent
\textbf{\emph{Total mass in stars.}} The masses of Pop~III stars are reduced by X-ray irradiation (see Paper~I). The physics is simple: in an X-ray background, the molecular fraction is higher and the gas temperature lower; this reduces the accretion rate ($dM/dt \propto M_J/t_{ff} \propto c_s^3$) onto the protostar and its final mass. As seen in the middle panels of Fig.~\ref{fig:massc}, the initial accretion rates ($\sim 10^{-1}$ and $10^{-2}~\msyr$) are lower than that in Fig.~\ref{fig:massa} ($\gtrsim 10^{-1}~\msyr$). The total mass in Pop~III stars in three X-ray simulations without LW background (Run~B, Run~D, and Run~E) decreases with increasing X-ray intensity ($183, 156$, and $105~\msun$, see Table~\ref{tab:sim}). We also confirm the \HII and photodissociation regions are narrower and the outflows are slower with increasing X-ray irradiations due to the lower masses of the stars (e.g., $\sim 70~\kms$ in Run~E). The total final mass in Run~C, on the other hand, is slightly greater than that of Run~A due to the delayed collapse induced by the LW background ($579~\msun$). All simulations end with $\dot{M} \lesssim 10^{-4}~\msyr$ and thus we expect $M_{final}$ does not change significantly after that.

Accretion peaks exist in Run~C as in the previous case (see Fig.~\ref{fig:massa}). From $t \sim 20$~kyr to $40$~kyr, the accretion rate of S02 increases periodically (see middle panel of Fig.~\ref{fig:massc}) with the peaks coincident with the minimum separations in the binary. These peaks, however, are less pronounced than in Run~A due to the less eccentric orbit and the larger average separation ($\sim 2000$~AU). Periodic accretion is less pronounced in the other three simulations with X-ray irradiation because the gas discs are smaller in mass.

\noindent
\textbf{\emph{Outward migration.}} The stars migrate outward in all simulations, and we show examples in the bottom panels of Fig.~\ref{fig:massc}. S01 and S02 in Run~C orbit each other with an initial separation of $\sim 500$~AU. The distances from CoM increase up to $\sim 2000$~AU by $t \sim 90$~kyr. Due to the formation of the hierarchical binary, two young stars (yellow and blue) are orbiting their companions with short periods in relatively close orbits (separations $\sim 500$~AU). Due to the low multiplicity, the orbits in Run~E are simple. The stars' orbits expand with time as in Run~C; they are born on an eccentric orbit, but they eventually circularise. 

In Paper~II we found that outward migration tends to be slower and to lower distances in simulation with increasing X-ray irradiation. This trend is not observed in this work: the maximum separation in Run~E ($J_{X0,21}=10^{-2}$) is larger than in Run~C ($J_{X0,21}=10^{-4}$). We simulated the system for a relatively short time for Run~B and Run~D (since accretion is suppressed early), but we find their orbits also expand with time. We will further explore the physical explanation for this ubiquitous outward migration in an upcoming paper.

\subsection{Results for a smaller mass halo}
Halo~3 has a lower mass than Halo~2 at all redshifts  (see Paper~I), hence Pop~III star formation in this halo is sensitive to lower values of the X-ray background, as shown by Fig.~\ref{fig:redshift}.

\noindent
\textbf{\emph{Multiplicity.}} Without an X-ray background, Halo~2 (Run~A) is dominated by three massive stars ($M\sim 200, 200$ and $100~\msun$) while Halo~3 (Run~F) has a less massive hierarchical quadruple system with masses between $M\sim 50~\msun$ and $70~\msun$. However, at a late time, Halo~3 becomes a more complex system as several smaller mass protostars form through late-time fragmentation ($N_{final}=7$, see Table~\ref{tab:sim}). Similarly to the case of Halo~2, the run with $J_{X0,21}=10^{-5}$\ in a LW background for Halo~3 (Run~G), has the largest multiplicity ($N_{final} = 10$) among the runs with RFB (see Fig.~\ref{fig:massg}). This is probably due to the delayed redshift of formation of the Pop~III stars and therefore a higher mass halo  when irradiated by a strong LW background. However, the trend of decreasing multiplicity with increasing X-ray background is also found in Halo~3: for  $J_{X0,21}=10^{-4}$ (Run~H) only two stars form in an eccentric orbit ($e \approx 0.5$).
\begin{figure}
    \centering
	\includegraphics[width=0.48\textwidth]{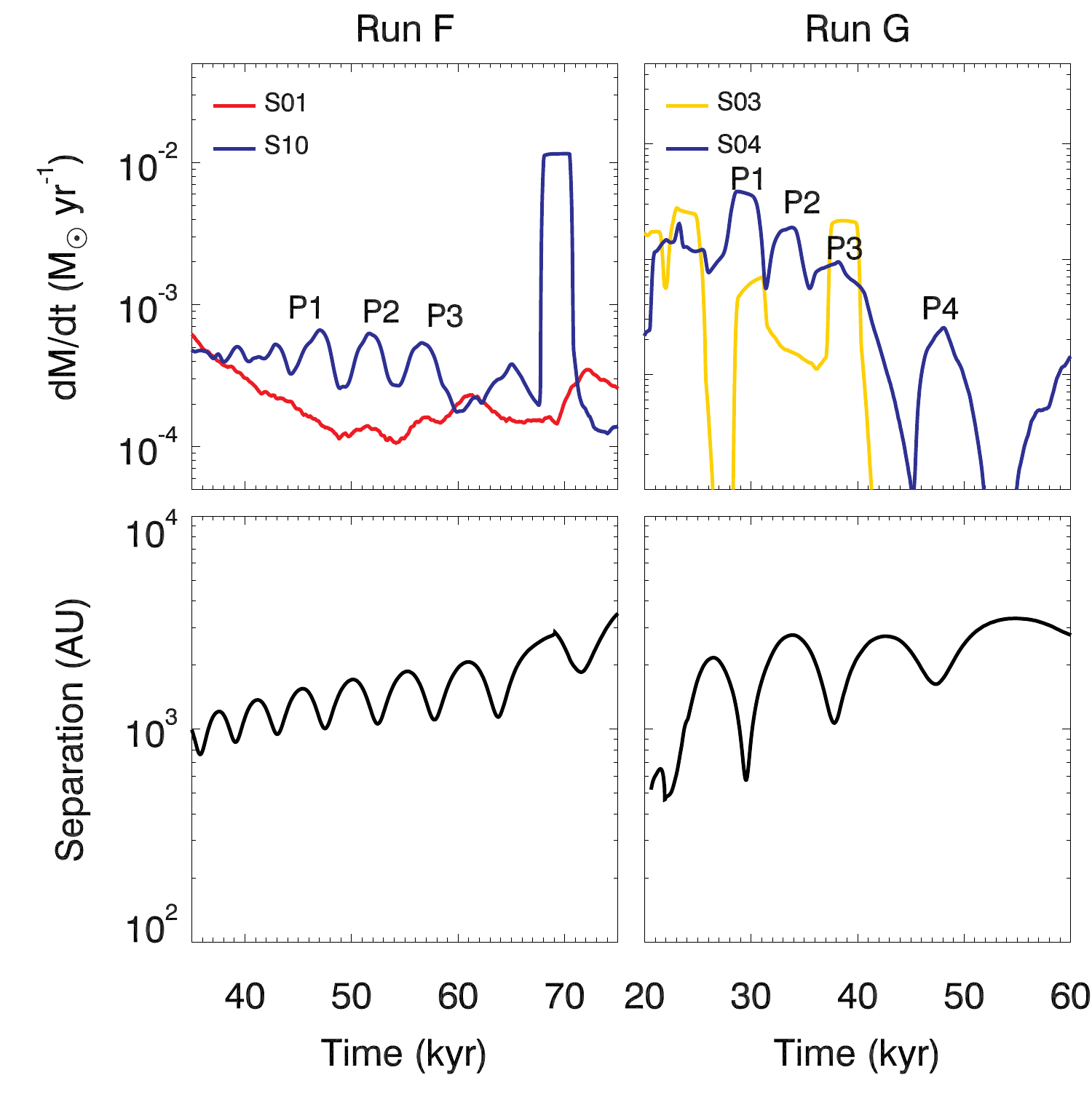}
    \caption{Top left: Accretion rates of S01-S10 binary in Run~F. We highlight prominent peaks with P1, P2 and P3. Bottom left: Separation between the same two stars. Top right: Accretion rates of S03-S04 binary in Run~G. Clear accretion peaks of S04 are marked with P1, P2, P3 and P4. Bottom right: The peaks except P2 coincide with the minimum separation and thus have a gravitational origin. P2 takes place at the apocenter and this is due to the accretion of the gas in a spiral arm.}
    \label{fig:pacc_simf}
\end{figure}

\noindent
\textbf{\emph{Total mass in stars.}}
The protostars in Halo~3 (Run~F) are less massive than in Halo~2, producing weaker RFB. This creates narrower bipolar \HII and photodissociation regions, thus suppression of the accretion rate is less significant. As it can be observed in the middle panels of Fig.~\ref{fig:massg}, the total accretion rate remains nearly constant as a function of time, although at a lower level than in the case without feedback (Run~F/noFB). The accretion rates of individual stars are very low ($\sim 10^{-4}~\msyr$) but the total accretion rate is about one order of magnitude higher as the multiplicity becomes larger at late times. In addition to the slow gas accretion, mergers between stars also contribute to the mass growth of individual stars. The mass of S10 after the merger is greater than the sum of the two masses because of the enhanced gas accretion triggered by tidal effects on the mini-discs. 

The trend with X-ray intensity is the same as for Halo~2: without X-ray irradiation (Run~F) the total mass is $338~\msun$, while for $J_{X,21}=10^{-4}$ (Run~H) the total mass is $203~\msun$. As in Halo~2, the case with LW irradiation and weak X-ray background (Run~G) has a greater total mass ($539~\msun$) due to the delay formation.

The protostars in Halo~3 also grow periodically as in Run~A. To better demonstrate the physical mechanism,  in Fig.~\ref{fig:pacc_simf} we plot the accretion rates and separations of selected binaries in Run~F and Run~G. For Run~F, the accretion rate of S10 (blue) shows multiple peaks with the clearest ones between $t \sim 40$~kyr and $60$~kyr. We marked them with P1, P2 and P3 in the top left panel. They have the same period and are in phase as the orbital separation: the peak accretion happens at the minimum separations in an eccentric orbit (bottom left). We thus conclude that the accretion is modulated by the tidal interaction of the mini-discs as discussed in Section~\ref{sec:sima}. This periodic behaviour, however, is less pronounced for S01 (red) because the companion (S10) and its circumstellar disc are smaller in mass and therefore the tidal force on the S01 disc is weaker. The other binary (S03-S05) in the same simulation evolves similarly. Binary stars in Run~G behave similarly and an example is shown in the right panels. Three peaks (P1, P3 and P4) take place at the minimum separations meaning the sink growth is driven by the interaction between discs. Interestingly, P2 happens at an apocenter of the orbit. At this time S04 passes through a spiral arm feature and therefore accretes the gas in it. As aforementioned, a boost of the accretion onto the bigger star (S03) is not clear because of the low mass of the companion (S04). Also in Run~H, the interaction in an eccentric binary boosts the gas accretion periodically, however, the periodic accretion does not last long as the accretion rate drops to $\sim 10^{-6}~\msyr$ in $\sim 40$~kyr.
\begin{figure}
    \centering
	\includegraphics[width=0.48\textwidth]{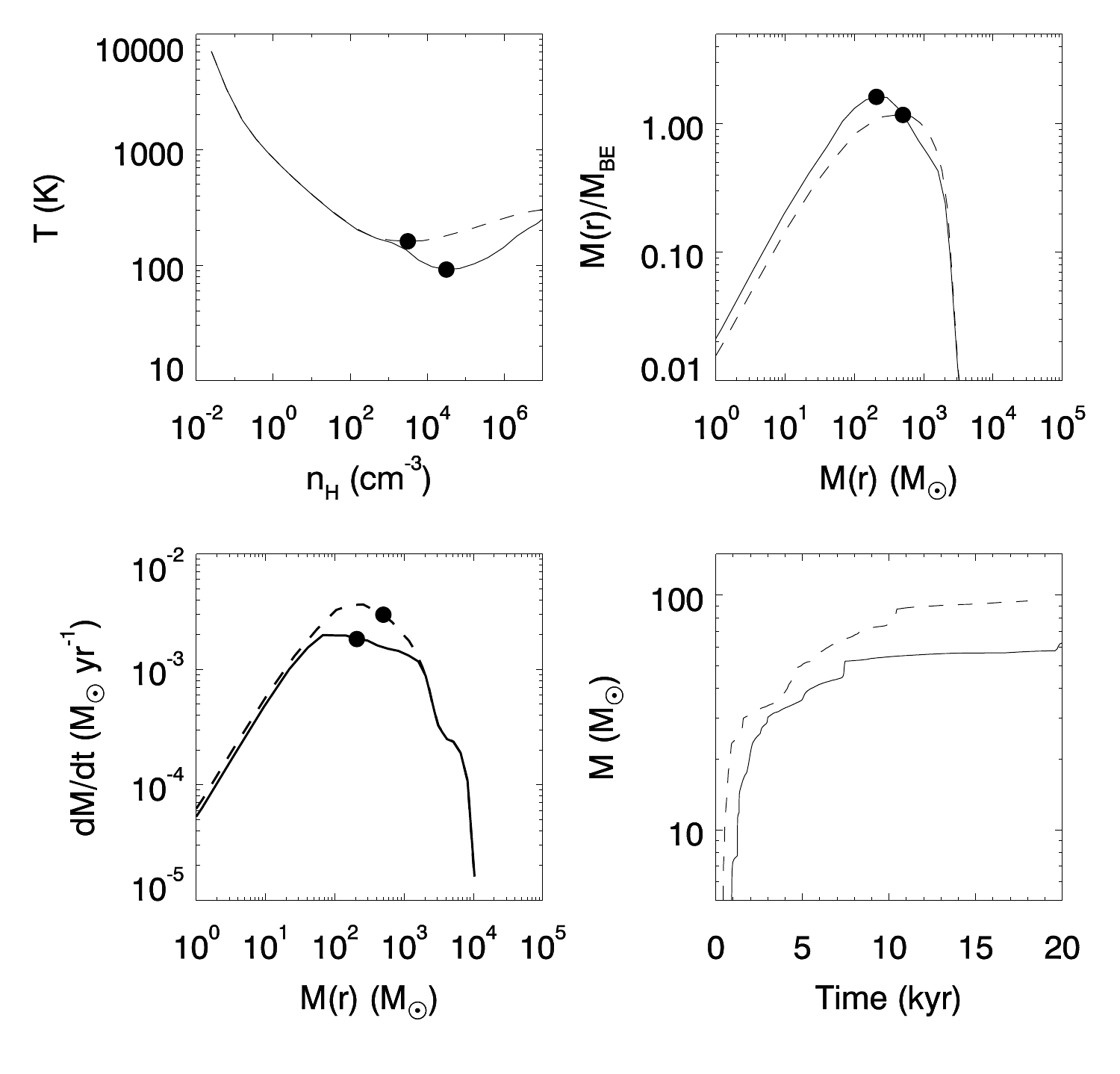}
    \caption{Top left: Phase diagram of Run~E (solid line) and Run~E_noHD (dashed line). In the former, the temperature drops to 92~K thanks to HD cooling. Top right: The ratio of the enclosed mass ($M(r)$) to Bonner-Ebert mass \citep[$M_{BE}$,][]{ebert1955,bonnor1956}. The peak radius is defined as the characteristic radius of the cloud. Due to the additional cooling in Run~E the cloud is smaller in size. Bottom left: Accretion rate. The smaller size results in a lower accretion rate. Bottom right: Total mass in sinks. Due to the reduced cloud size the sinks in Run~E are smaller in mass.}
    \label{fig:HD}
\end{figure}

\noindent
\textbf{\emph{Outward migration.}}
As in Halo~2, most of the stars migrate outward in Halo~3. One exception is S11 in Run~D (magenta, the bottom panel of Fig.~\ref{fig:massg}). Forming at $2 \times 10^4$~AU, it migrates inward and finally merges with S10 after 10~kyr from its formation. Another distinct feature of Halo~3 is star formation in the proximity of Lagrange points L4 and L5 around the main binary. Using hydrodynamics simulations, several authors \citep{lyra2009,montesinos2020} showed that gas and dust in protoplanetary discs accumulate at L4 and L5. 


\section{Effect of HD Cooling}
\label{sec:HD}
The effect of HD cooling kicks in at about $T \sim 150$~K. The temperature of primordial gas may not reach this threshold with \hm\ cooling alone (Paper~I). However, additional ionization by cosmic rays \citep{nakauchi2014}, X-rays (\citealp{jeon2014a,hummel2015}; Paper~I), or star formation in relic \HII regions \citep{RicottiGS:2001}, may enhance \hm\ abundance and reduce the gas temperature enough to trigger HD cooling and thereby reduce the typical Pop~III mass scale. As in this work, previous simulations with X-ray irradiation have taken into account HD formation and cooling \citep{jeon2014a,hummel2015}. However, in this section we quantify its importance with respect to the case in which HD is neglected.

In Fig.~\ref{fig:HD} we compare two simulations with strong X-ray irradiation in Halo~2 (Run~E), including (solid lines) and excluding (dashed lines) HD chemistry/cooling.
The phase diagrams ($n$ vs $T$, the top left panel) show that the minimum gas temperature, $T_{min}$, at $n \sim 10^4$~\cc\ is reduced by the HD cooling from $T_{min} = 161$~K to $92$~K. Because of the lower gas temperature, HD cooling reduces the size and mass of the collapsing core (top right panel) and reduces the gas accretion rate (bottom left). The bottom right panel compares the total mass in Pop~III stars in the simulations with and without HD. Since we run the case without HD cooling (Run~E\_noHD) for a relatively short time ($\sim 18$~kyrs), we only compare the masses at early times. The figure shows a reduction of $\sim 50$\% of the total mass in Pop~III stars, from $95~\msun$\ to $57~\msun$\ at 18~kyr.

HD cooling has a small effect also in simulations of Halo~3, however the decrease of $T_{min}$ and therefore the decrease of the total mass in stars is weak for $J_{X0,21} < 10^{-3}$. Note that $T_{min}$ may be as low as $\sim 100$~K and trigger HD cooling also in the absence of X-ray irradiation, depending on the collapse history of the halo (\citealp[e.g.][]{hirano2014}; HR15). However, an X-ray background has a systematic effect in reducing $T_{min}$ below 100~K, making HD cooling important in further reducing the temperature and therefore the masses of Pop~III stars.


\section{Discussion}
\label{sec:discussion}
\begin{figure}
    \centering
	\includegraphics[width=0.48\textwidth]{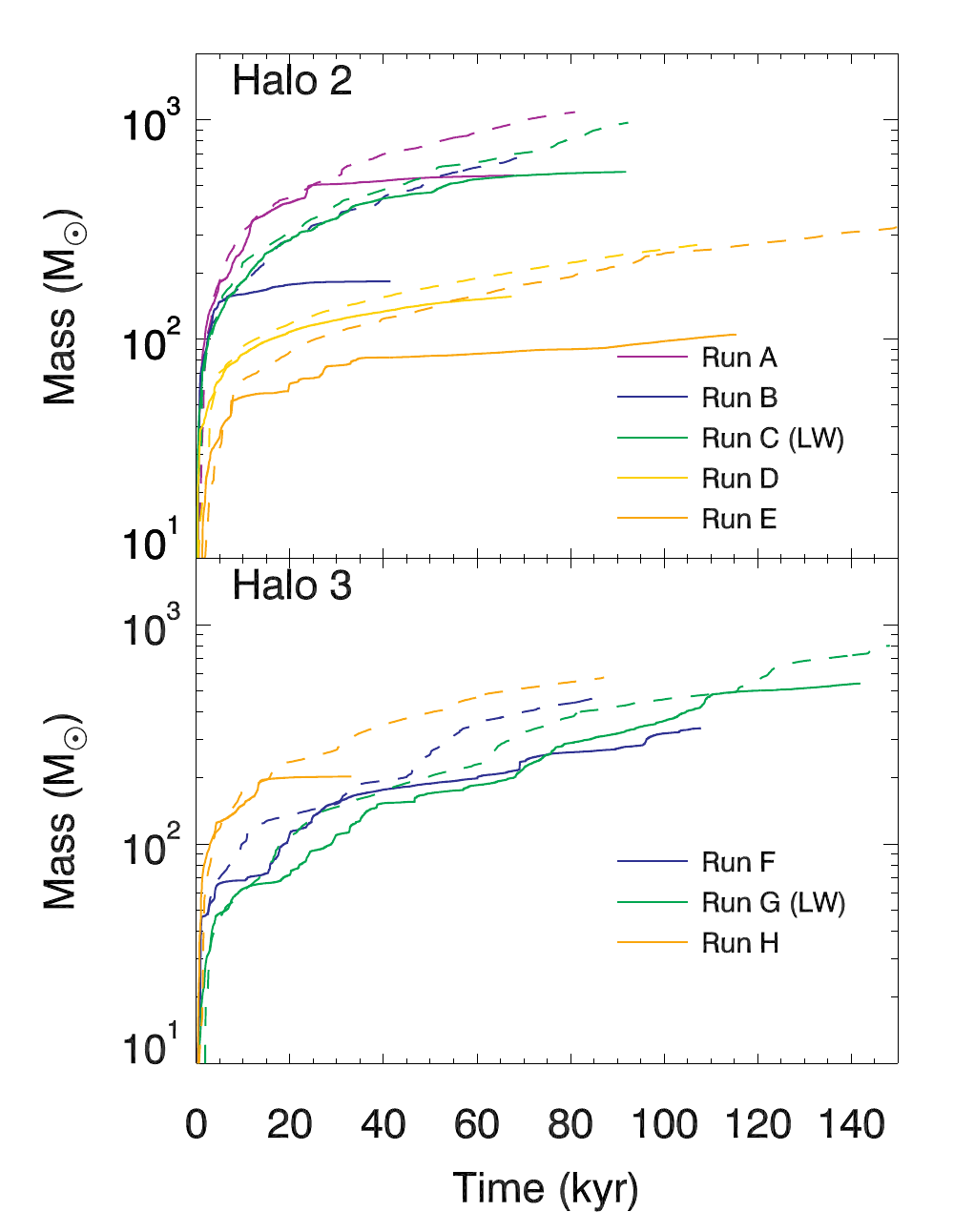}
    \caption{Total mass of protostars. Top and bottom panels show the results for Halo 2 and Halo 3, respectively. The simulations with RFB are shown with solid lines and those without RFB are shown with dashed lines. ``LW'' indicates the disc is irradiated by a strong LW background.}
    \label{fig:mass}
\end{figure}

\hide{
Some authors \citep{mckee2008,hosokawa2011} studied the role of protostellar FB in suppressing Pop~III star growth. To extend this understanding we perform RHD simulations of Pop~III stars and primordial discs irradiated by LW+X-ray backgrounds. In different haloes and radiation backgrounds, the mass of Pop~III stars is reduced by RFB as shown in Fig.~\ref{fig:mass}. Note that the masses in the fiducial runs with RFB (solid lines) are lower than the noFB counterparts (dashed lines) confirming the earlier results are valid in various environments.}
\begin{figure}
    \centering
	\includegraphics[width=0.48\textwidth]{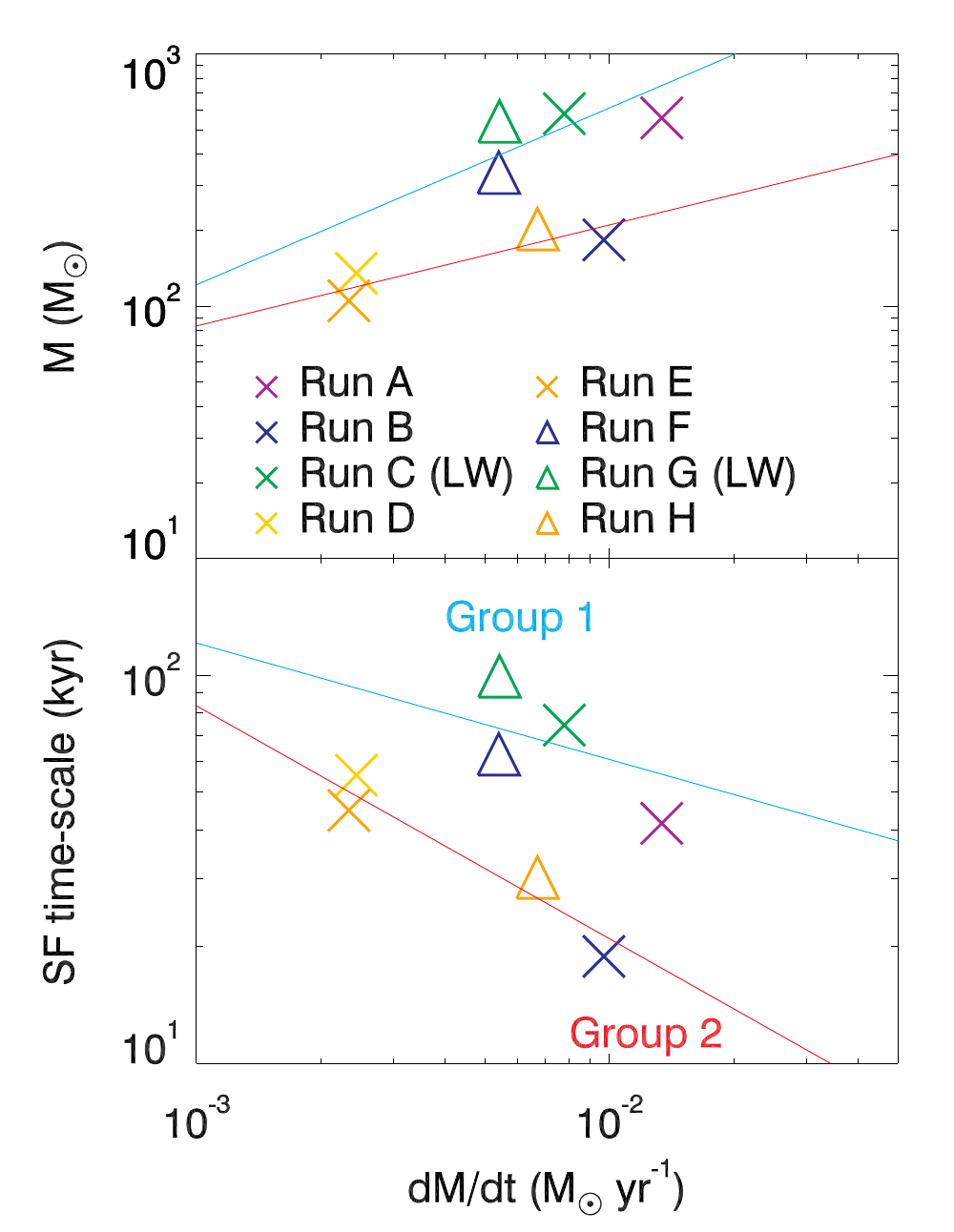}
    \caption{The final mass (top) and star formation timescale (bottom) in the fiducial simulations as a function of accretion rate. We categorize the simulations into two groups: Group~1 and Group~2. Group~1 (cyan) includes runs with no X-ray background and runs with non-zero LW background and weak X-rays: Run~A, Run~C, Run~F, and Run~G. On the other hand, Group~2 (red) consists of X-ray-only cases. The cyan line is the empirical relationship in HR15 and Group~1 follows this trend well. The red line is the fit to Group~2. SF timescale is defined as $M_{final}/({\rm d}M/{\rm d}t)$ with ${\rm d}M/{\rm d}t$ is defined by $M_{final}/t_{final}$ derived in the noFB counterpart of each simulation (see the text for detail). As in Fig.~\ref{fig:mass}, simulations with LW backgrounds are shown with label ``LW''.}
    \label{fig:formula}
\end{figure}
\textbf{\emph{Accretion timescale in X-ray background}}. To quantify the efficiency of RFB, we can estimate the star formation timescale, $\tau_{SF} \equiv M_{pop3,tot}/(dM/dt)$, defined as the final mass of Pop~III stars over the mass accretion rate. According to equation~(\ref{eq:tsf}), $\tau_{SF}$ is shorter for a higher accretion rate, making the final mass of Pop~III stars less sensitive to the accretion rate ({i.e.}, $M_{final} \propto (dM/dt)^\alpha$ with $\alpha<1$). In other words, Pop~III protostars are more massive in a rapidly accreting cloud, but their stronger RFB shuts down accretion more rapidly. In our simulations we include RFB, hence we can directly derive  $\tau_{SF}$ and compare our results to HR15. The caveat is that our statistical sample is small: we only have eight simulations in two halos.

In Fig.~\ref{fig:mass} we show the total mass in Pop~III stars including and excluding RFB for Halo~2 and Halo~3 in different X-rays and LW backgrounds. Because $dM/dt$ and the masses of Pop~III stars decrease when increasing the X-ray background, following the aforementioned logic, we also expect a monotonic increase of $\tau_{SF}$. However, this is not observed in all our simulations, suggesting that $\tau_{SF}$ may be determined by another physical parameter in addition to the strength of RFB.
\begin{figure}
    \centering
	\includegraphics[width=0.48\textwidth]{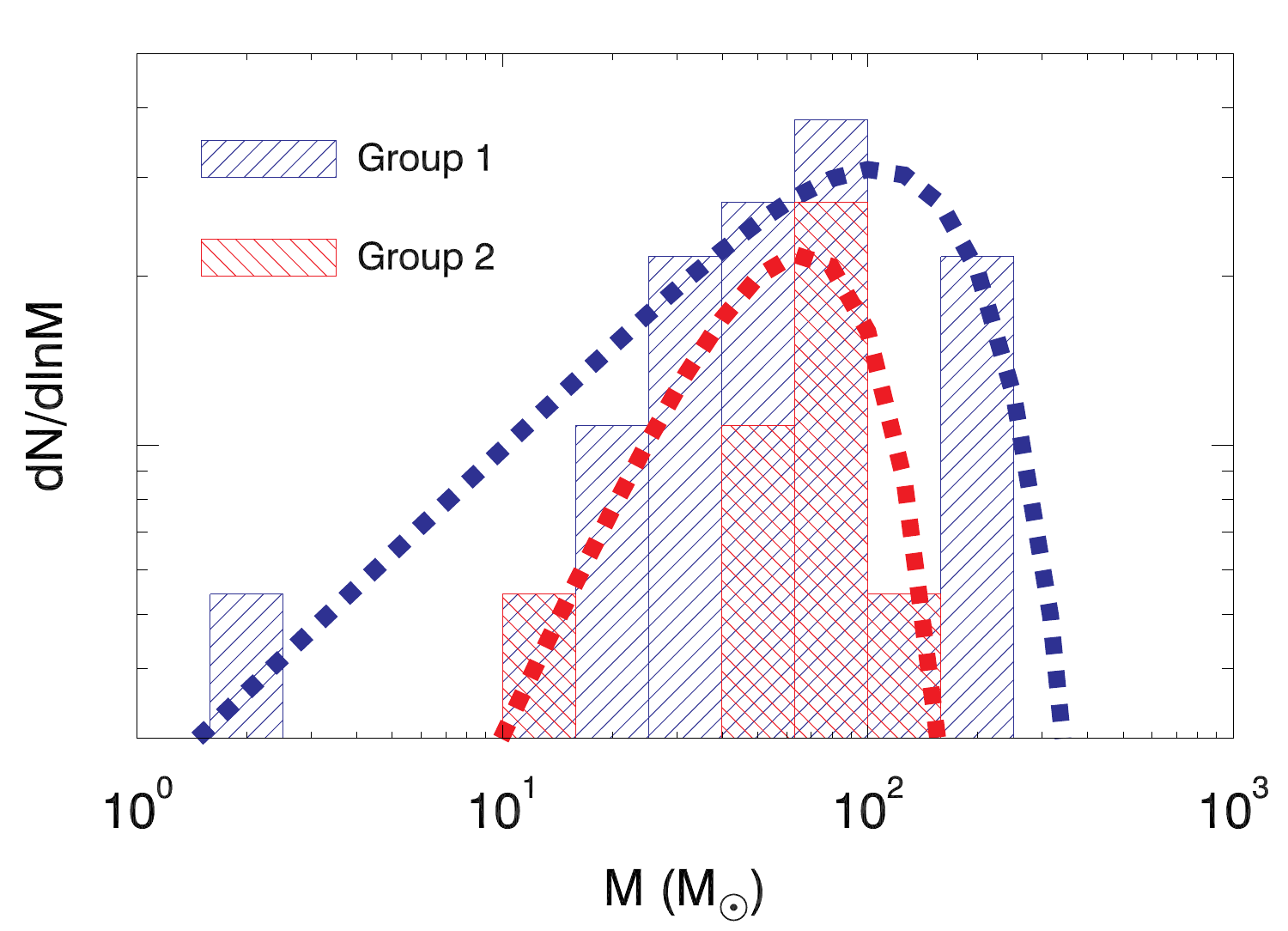}
    \caption{The IMFs for the runs in Group~1 with weak or no X-ray background (shaded blue histogram) and Group~2 with stronger X-ray irradiation (red shaded histogram). We count only Pop~III stars that survive without merging in the last snapshots. We also show the best fits (dashed lines) using equation~(\ref{eq:IMF}).}
    \label{fig:imf}
\end{figure}
\begin{figure*}
    \centering
	\includegraphics[width=0.95\textwidth]{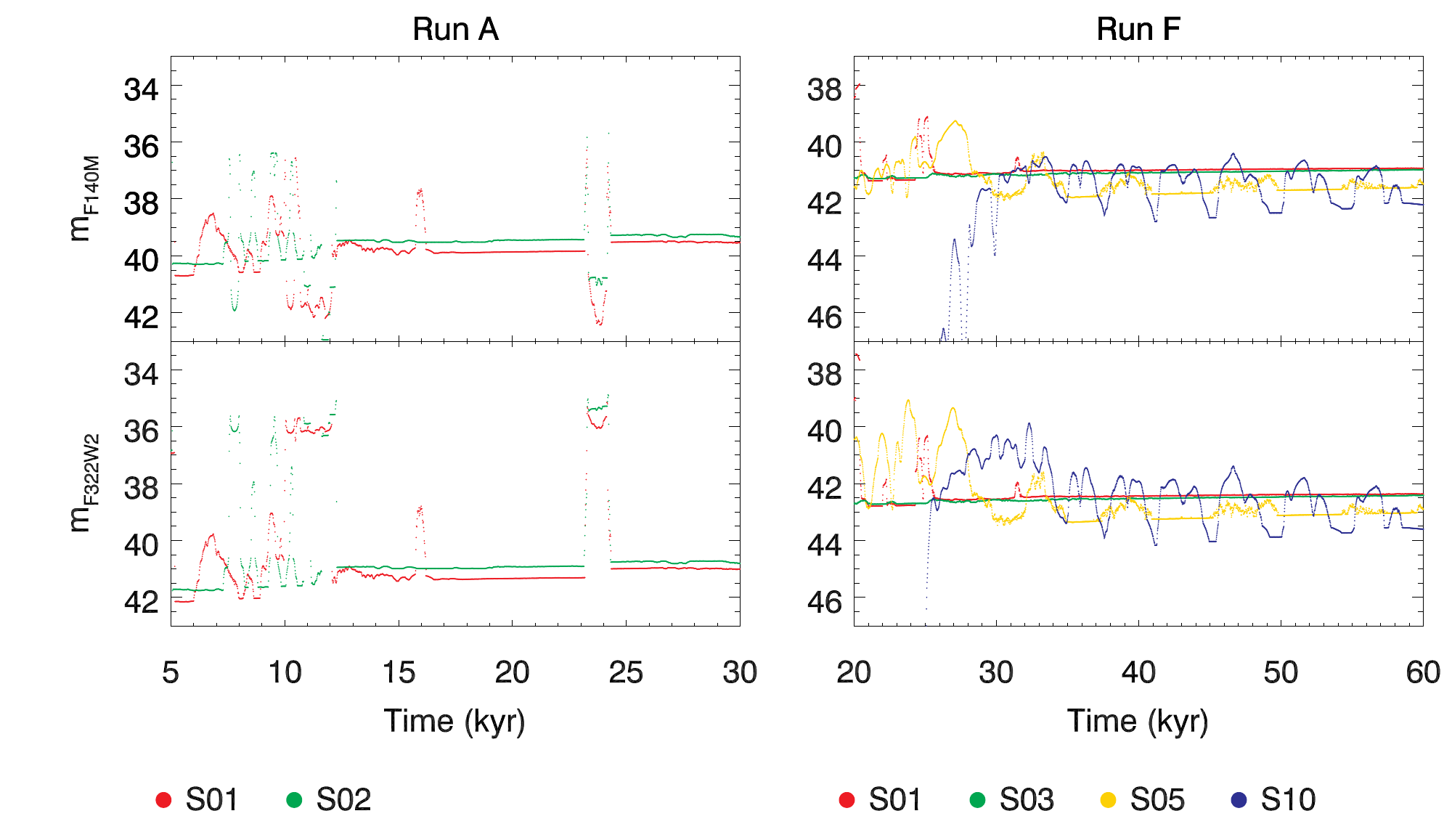}
    \caption{Apparent magnitudes of protostars in two {\it JWST} NIRCam bands: F140M and F322W2. We assume the protostars emit like a perfect blackbody and they are at $z=6$. With this redshift, the centres of the bands trace the rest-frame non-ionizing UV ($\sim 2000$~\AA) and visible light ($\sim 4600$~\AA). The left panels show S01-S02 binary in Run~A and the right panels show the stars in the hierarchical binary in Run~F. In the latter, S01 and S10 are in a close binary and S03-S05 are in the other close binary.}
    \label{fig:magnitude}
\end{figure*}

To be more quantitative in Fig.~\ref{fig:formula} we plot $M_{final}$ and $\tau_{SF}\equiv M_{final}/(dM/dt)$ as a function of $dM/dt$ estimated from the last snapshot for the noFB simulations: i.e., $dM/dt \equiv M_{final}/t_{final}$. We find that $dM/dt$ estimated from the noRF simulations agrees within a small scatter with the accretion rate at the characteristic radius ($dM/dt|_{cr}$) as defined in HR15. The figure shows that $M_{final}$ and $\tau_{SF}$ in the X-ray-only (without LW background) simulations are systematically lower with respect to the empirical relationships in HR15, shown by the solid blue lines. For this reason, we categorize the simulations into two groups. Group~1 (Run~A, Run~C, Run~F and Run~G) consists of zero/weak X-rays and the two simulations including the LW background, which follow the relationship in HR15 (blue line). Group~2 (Run~B, Run~D, Run~E and Run~H) consisting of simulations with moderate to high X-ray irradiation (J$_{X0,21} \geq 10^{-4}$), showing significant discrepancy from the HR15 relationships. The solid red lines are power-law fits to the data points in Group~2:
\begin{equation}
    M_{final} = 126~\msun \left( \dfrac{{\rm d}M/{\rm d}t}{2.8\times10^{-3}~\msyr} \right)^{0.4},
\label{eq:formula}
\end{equation}
and
\begin{equation}
    \tau_{SF} = 45~{\rm kyrs} \left( \dfrac{{\rm d}M/{\rm d}t}{2.8\times10^{-3}~\msyr} \right)^{-0.6}.
\label{eq:tsf2}
\end{equation}
We interpret the lower Pop~III masses and shorter star formation timescales found for Group~2, as due to the effect of X-rays in reducing the typical masses of protostellar discs (see Paper~I). X-rays increase H$_2$ formation and cooling at high densities, reducing the accretion rate of the collapsing core, but also heats the IGM and the low-density gas at the outskirt of the halo, thereby somewhat decreasing the baryon fraction ($f_b$) in the minihalo. For example, when the central density is $10^7$~\hcc\ (Fig.~\ref{fig:redshift}), $f_b$ is close to the cosmic average ($\Omega_{b}/\Omega_{DM} \sim 0.2$) if J$_{X0,21} \lesssim 10^{-4}$ but it is a factor of 2 to 4 smaller in simulations with  J$_{X0,21} \gtrsim 10^{-2}$, depending on the halo mass. Indeed in our simulations we observe smaller and less massive discs/cores with strong X-ray irradiation (see also Fig.~6 in Paper~II), even when keeping constant $dM/dt|_{cr}$ at the critical radius and therefore the total mass in Pop~III stars (see Fig.~\ref{fig:ic} and Paper~I). The smaller discs are more fragile to RFB, therefore the mass growth is suppressed more rapidly even in presence of a weak RFB from lower-mass protostars.

\noindent
\textbf{\emph{Initial mass function of Pop~III stars.}}
In Paper~II we found that the IMF of Pop~III stars can be described as a power-law with an exponential cutoff at high mass:
\begin{equation}
    \frac{{\rm d}N}{ {\rm d} \ln M} = A M^\alpha \exp{\left[-\left(\frac{M}{M_{cut}}\right)^2\right]},
    \label{eq:IMF}
\end{equation}
where we find $\alpha=0.49, 1.53$ and $M_{cut}=229,61~\msun$ for weak and strong X-ray irradiation, respectively.
In strong X-rays, the IMF has a lower cutoff mass $M_{cut}$ and steeper power-law slope $\alpha$ because the gas cloud becomes less massive and fragmentation is suppressed. In this work we find results roughly in agreement with Paper~II as shown in Fig.~\ref{fig:imf} with fitting parameters given in Table~\ref{tab:imf}. For comparison, in their galaxy formation simulation \citet{wise2012} assumed a Pop~III MF with the peak at $\sim 100~\msun$, while \citet{susa2014} found an MF with a peak at a lower mass (a few $\times~10~\msun$). The MF in \citet{stacy2013}, on the other hand, is flat without a peak. 

\begin{table}
    \caption{Parameters of equation~(\ref{eq:IMF}).}
    \centering
    \begin{threeparttable}
    	\label{tab:imf}
    	\begin{tabular}{ | c | c | c | c | c |}
        \hline
        \multicolumn{2}{l|}{Paper~II} & & & \\
        \hline
		X-ray & $\alpha$ & $M_{cut}$ & $A$ & $M_{peak}$\tnote{a} \\
		\hline
		Weak & 0.490 & $229~\msun$ & 0.169 & $113~\msun$  \\
		\hline
		Strong\tnote{b} & 1.53 & $61~\msun$ & 0.00692 & $53.4~\msun$   \\
		\hline
		\multicolumn{2}{l|}{This work} & & & \\
		\hline
		FB & $\alpha$ & $M_{cut}$ & $A$ & $M_{peak}$ \\
		\hline
		Group~1 & 0.620 & $188~\msun$ & 0.237 & $105~\msun$  \\%
		\hline
		Group~2\tnote{c} & 1.41 & $79~\msun$ & 0.012 & $66~\msun$  \\%
        \hline
	    \end{tabular}
	    \begin{tablenotes}
	        \item[a] $M_{peak} = M_{cut} \sqrt{\alpha/2}$.
	        \item[b] J$_{X0,21} \geq 10^{-3}$.
	        \item[c] J$_{X0,21} \geq 10^{-4}$ and include no LW simulations.
        \end{tablenotes}
    \end{threeparttable}
\end{table}

\noindent
\textbf{\emph{Protostellar variability.}}
As discussed in Section~\ref{sec:halo2}, Pop~III protostars grow periodically via enhanced gas accretion in an eccentric orbit. As the luminosity is a function of mass and accretion rate, Pop~III protostars may display variability of their luminosity and effective temperature potentially on interestingly short timescales, perhaps observable by {\it JWST} or {\it NRST}. 

As shown in Fig.~\ref{fig:magnitude}, individual protostars of about 100~M$_\odot$ cannot be directly detected by {\it JWST}, unless their luminosity is magnified by a factor $\sim 100$ by a gravitational lens. For instance, in the rest-frame optical filter $m_{F322W2}$ in Run~A (bottom left panel) the peak magnitude is $\sim 34-35$, below {\it JWST} detection limit ($\sim 30-31$). The periodic increase in accretion can bring the star from being bright in the UV band (effective temperatures $T_{eff} \sim 10^5$~K) to a red-supergiant phase with peak brightness in the optical bands and faint in the UV \citep{hosokawa2009,hosokawa2010,hosokawa2016}.
Due to the short mean free path of UV photons, typically Pop~III stars can be identified at high-z by nebular line emission: \HeII recombination lines and the large equivalent width of the Ly~$\alpha$ line. However, during the strong accretion phase, Pop~III stars shine in the rest-frame visible at nearly the Eddington luminosity and in principle the light of the star can be observed directly. As a reference, a Pop~III protostar of mass $M$ at $z=6$ has magnitude $m_{AB}\sim 34-2.5\log_{10}(M/100~M_\odot)$ in the rest frame visible when the accretion rate is above the critical threshold of $10^{-2}$~M$_\odot$~yr$^{-1}$. However, the strongly accreting phases are shorter than the lifetime of the star, hence these objects are relatively rare.

During the bloated phase, the star becomes brighter in red but fainter in blue due to the low $T_{eff}$, therefore the magnitude reaches $m_{F322W2} \sim 34$ in the rest frame optical. When observed with a narrower band filter in the rest frame UV (F140M, top left panel) the apparent magnitude actually increases (i.e., the star becomes fainter) during the strongly accreting phase. In Run~F (right panels) a low-mass star (S10) orbits its massive companion (01) and shows a clear variability. In general, S10 is fainter than S01 in rest frame UV due to the lower mass (it has a lower effective temperature and size), but becomes brighter periodically when is rapidly accreting (top right, Fig.~\ref{fig:magnitude}). On the contrary, S10 is brighter than the companion in the rest-frame optical bands most of the time but still shows periodicity on relatively short timescales (bottom right).

The periodic modulation of the luminosity due to pericenter passages depends on the total mass of the binary and separation: in our simulations it varies on timescales between several tens of kyr to a few 100 years. However this is simply because close binaries with separation $<200$~AU are not resolved in our simulations, but we expect that harder Pop~III binaries are common as we find several protostars that artificially merge with each other due to our sink prescription (see Section~\ref{app:merger}). In addition, as it can be observed in Fig.~\ref{fig:magnitude} (left), the magnitude of the star as it transitions to the supergiant phase can decrease by 7 magnitudes (about a factor of 600 increase in luminosity) in a few tens of years. Even considering time dilation at $z=6$, such a rapid increase can be observed with a short time baseline. Note that the star may become bloated on a very short timescale but return to its normal size on a longer timescale (comparable to the Kelvin-Helmholtz timescale), not modelled here.

\noindent
\textbf{\emph{Possible channels for GW sources.}}
Close Pop~III binaries, given their large masses, are likely to produce IMBH binaries that my be sources of gravitational waves detectable by LIGO/VIRGO. Unfortunately, due to resolution limitations in our simulations we cannot directly resolve the orbits of binary stars with separations $<200$~AU, even though the resolution of the simulations is about $13$~AU. 
This is because, stars are represented by sink particles that have radii of 8 cells (about $101$~AU), and we assumed that any two sinks with separation $< 202$~AU merge into one. To understand the behaviour of two stars closer than this threshold, we performed a test simulation dropping this assumption (see Appendix~\ref{app:merger}). In this test simulation the stars form close binaries with separations of $\sim 30$~AU and $80$~AU. Hence, we expect that such binaries are a common occurrence in Pop~III star systems and once they evolve to binary black holes (BBHs) with a smaller separation, they can possibly merge and emit detectable GWs \citep{liu2020}. It is beyond the scope of this paper to make predictions on the GW emission rate from such systems.

\begin{figure}
    \centering
	\includegraphics[width=0.48\textwidth]{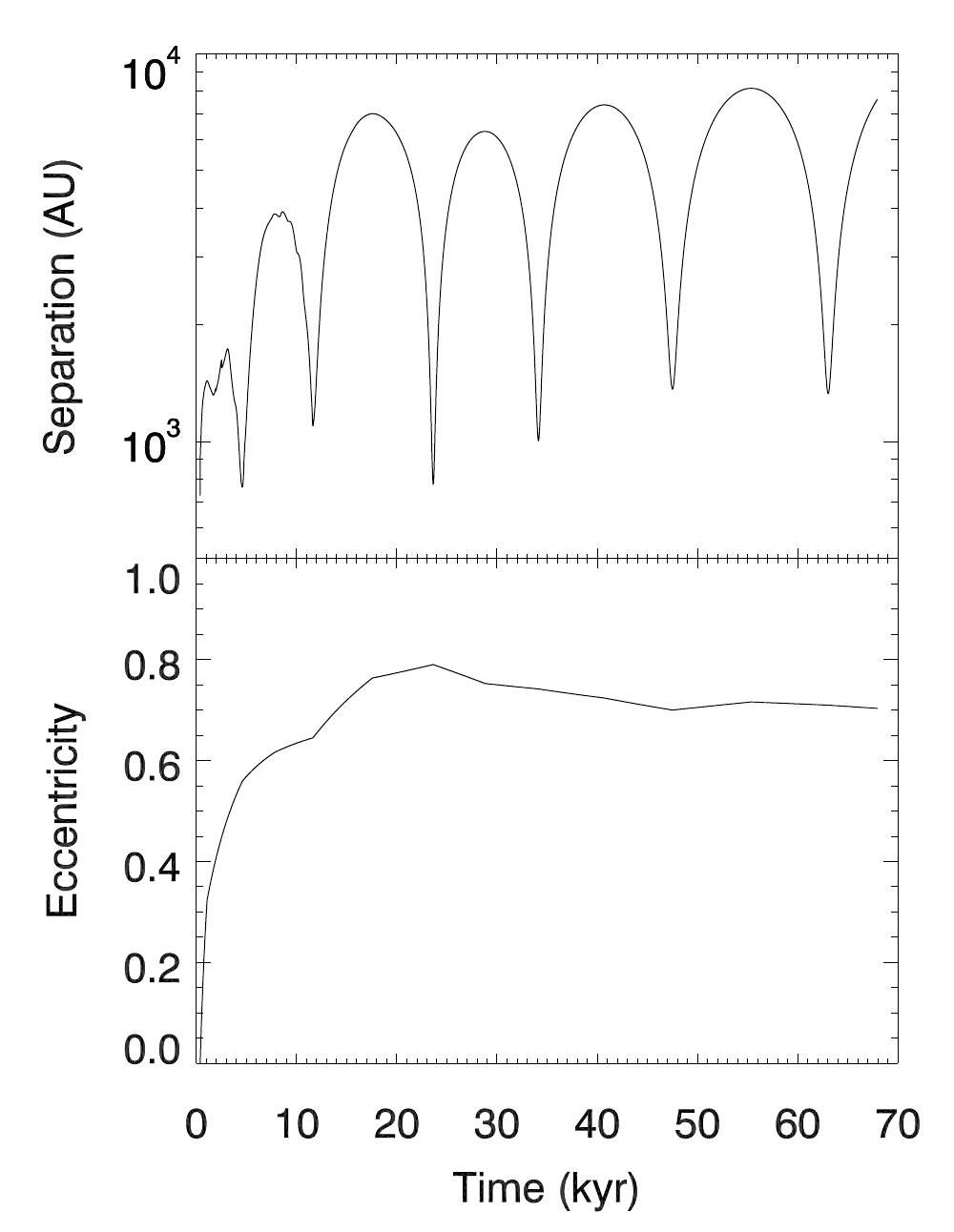}
    \caption{Top: Separation between S01 and S02 in Run~A. The minimum and maximum separations are $\sim 1000$~AU and $8000$~AU. Bottom: Eccentricity of the orbit.}
    \label{fig:ecc}
\end{figure}

Wide Pop~III star binaries, that are very common in our simulations, can also be potential GW sources. The physical mechanism allowing this channel of GW emission is dynamical hardening by other stars \citep{liu2020} and BBH mergers in wide eccentric orbits triggered by orbital excitation by field stars \citep{michaely2019,michaely2020}. When a wide eccentric orbit becomes more eccentric due to an encounter with a field star, the orbit may shrink due to the GW emission at each pericenter, allowing the BBHs to merge within a Hubble time. Several of the binaries in our simulations have rather circular orbits with eccentricities $e \lesssim 0.2-0.3$. But in our small sample of simulations we found some binaries with large eccentricities, hence such cases are not rare. For instance, the S01-S02 binary in Run~A starts with a large eccentricity ($e \sim 0.8$) as shown in Fig.~\ref{fig:ecc}, and later $e$ slightly decreases to $\sim 0.7$ but we expect it does not decrease further because there is no gas disc or Pop~III stars to circularise the orbit. Although its eccentricity and semi-major axis ($a \sim 5000$~AU) are lower than the values for the maximum merger probability of $e$ approaching unity and $a \sim 10,000$~AU found by \citet{michaely2019} (see their Fig.~1 and Table~1), the probability of merger is only a factor of 2 lower than the maximum value reported by \citet{michaely2019} in terms of the semi-major axis. However, the merger rate increases significantly for values of $e$ closer to unity than $e=0.9$. Nevertheless, we point out that this channel of GW emission can be important for Pop~III star systems given the ubiquitous outward migration and large eccentricities of the hierarchical binaries found in our simulations. Clearly, more work is needed to estimate IMBH merger rates at $z=0$ through this channel, and compare it to hard-binary merger rates.

\section{Summary}
\label{sec:sum}
Using radiative hydrodynamics simulations we explore the evolution of Pop~III protostars regulated by protostellar radiative feedback. The initial conditions for star formation are extracted from cosmological zoom-in simulations of two minihaloes in different X-ray and LW radiation backgrounds. In this paper, the third in a series, we implement sink particles, gas accretion and photon injection to model radiative feedback from protostars. Below we list the key findings and implications of this study.
\begin{enumerate}
    \item Protostellar feedback suppresses the growth of protostars as in earlier works \citep{hosokawa2011}. LW photons photodissociate \hm\ and increase the gas temperature to a few $10^3$~K in broad bipolar regions above and below the disc. Ionizing photons are initially trapped in the vicinity of the stars but eventually also produce narrower bipolar regions where the ionized gas reaches temperatures between $\sim 10^4-10^5$~K. Hot gases create an outflow and suppress the growth of protostars by reversing the gas flow. Feedback has instead negligible effect on stellar multiplicity during the first 100~kyrs.
    
    \item  Nearly all Pop~III star systems are hierarchical, i.e., binaries of binaries. Most commonly two equal-mass stars form near the centre of the protostellar disc and migrate outward. Around these stars two mini-discs become Toomre unstable and fragment forming binaries that also migrate outward. Later on, stars may form at the Lagrange points L4 and L5 of the system. Afterward, star formation becomes more stochastic due to the large multiplicity.
    
    \item We show that low-mass Pop~III protostars can form when rapidly ejected from the disc. Although this star has mass $\sim 2~\msun$ this result suggests low-mass Pop~III star might exist and be detected in the local universe.
    
    \item Another notable result is that often the stars in the disc have eccentric orbits, leading to a periodic modulation of their accretion rates and their luminosities (on time scales that vary between a few tens of years and several tens of kyr). 

    \item Pop~III protostars can enter a supergiant phase being bright in the rest-frame optical bands but faint in the UV due to the periodic increase in accretion. During this phase, they can be observed with the {\it JWST} directly rather than by nebular line emission. At $z=6$, $100~\msun$ star has $m_{AB} \sim 34$ and this can be above the detection limit of the {\it JWST} if magnified by gravitational lensing.
    
    \item We confirm previous results that a moderate X-ray background reduces the minihalo mass threshold for star formation, thus increasing the number of minihalos forming Pop~III stars per unit volume, but this increase is mitigated by a reduction of the typical masses and multiplicities of the stars. The stellar mass function also agrees with previous results \citep{park2021b} and we confirm the outward migration of the stars within the protostellar disc for nearly all the stars.

    \item We confirm that HD cooling is important in reducing Pop~III star masses in a strong X-ray background.
    
    \item From our small set of simulations we derive a relationship for the typical duration of the accretion phase of protostars due to their radiative feedback. We find that this timescale depends not only on the gas accretion rate (and therefore total mass in Pop~III stars), but also on the protostellar disc mass, that is systematically smaller for strong X-ray irradiation.
        
    \item The common occurrence of eccentric Pop~III binaries with large separation ($e \approx 0.8$ and $a \sim 5000$~AU) observed in our simulations, implies that a new channel for GW emission -- binary black hole mergers through orbital excitation \citep{michaely2019,michaely2020} -- is likely for Pop~III systems. Therefore, Pop~III stars may contribute more than previously thought to gravitational wave signals detected by LIGO/VIRGO observatory. Calculation of detection rates through this channel is a topic of future work.
\end{enumerate}
\section*{Acknowledgements}

The authors thank Dr. Takashi Hosokawa for kindly sharing the radiative feedback model of protostars and Dr. Erez Michaely for insightful discussion. We made use of the Deepthought2 cluster operated by the University of Maryland (http://hpcc.umd.edu) to perform the all simulations in this work. This research is supported by Grants-in-Aid for Scientific Research (KS: 21K20373) from the Japan Society for the Promotion of Science and the Hakubi Project Funding of Kyoto University (KS). 

\section*{Data Availability}

The data underlying this article will be shared on reasonable request to the corresponding author.



\bibliographystyle{mnras}
\bibliography{reference} 



\appendix

\section{Radiative Transfer Recipes}

\subsection{Importance of X-ray Self-Shielding}
\label{app:shd}
\begin{figure}
    \centering
	\includegraphics[width=0.48\textwidth]{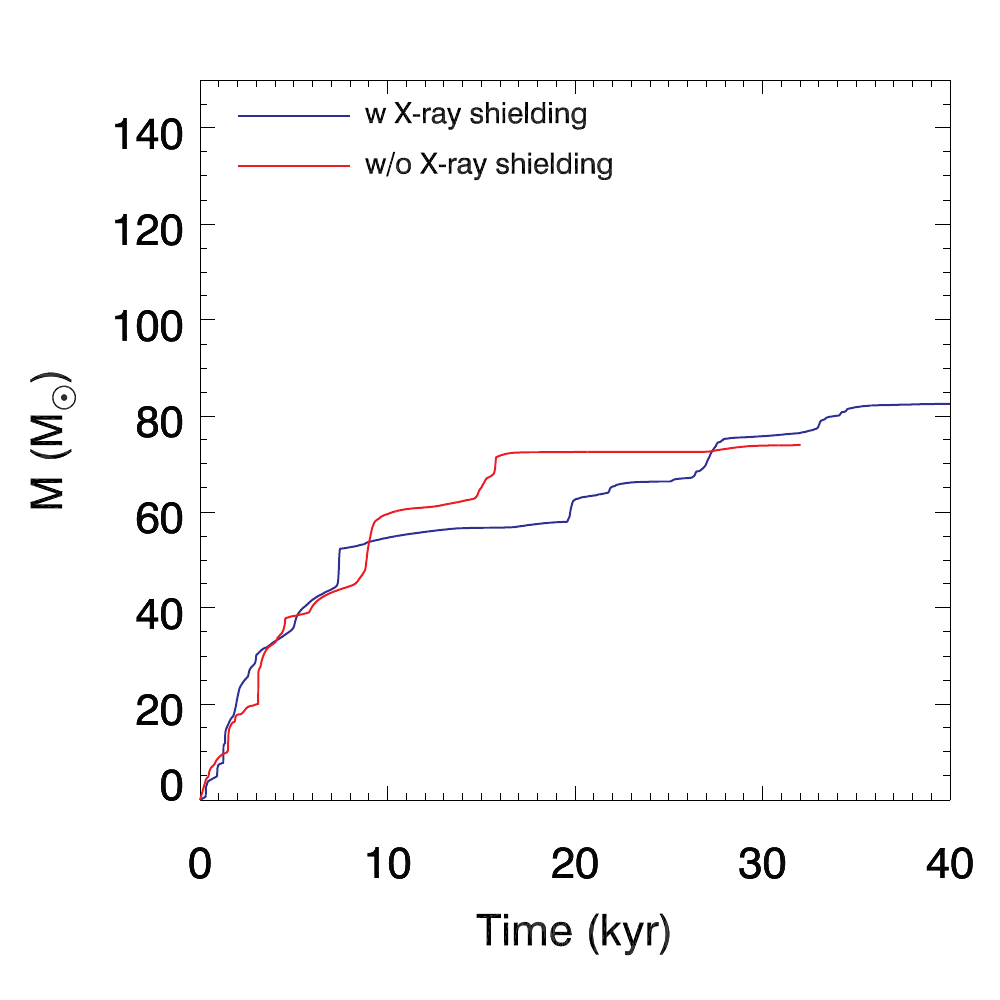}
    \caption{Total mass in Pop~III stars including (blue line) and excluding (red line) X-ray self-shielding for Run~E ($J_{X0,21}=10^{-2}$). All fiducial runs include self-shielding.}
    \label{fig:xray}
\end{figure}

The test results on the effects of X-ray self-shielding are presented in Fig.~\ref{fig:xray}. We ran a simulation with the same parameters as Run~E but without including X-ray self-shielding (shown in red). We find that the sink mass is larger between $t \sim 15$ and $25$~kyr when the disc is not shielded from the X-ray background. This is consistent with \citet{hummel2015} in that the cooling is more efficient without shielding. The difference, however, is small and the results converge by $t \sim 30$~kyr. This small difference can be explained in the following way. As discussed in Paper~I and Paper~II properties of the stars are determined by the mass of the hydrostatic core at $n_{H} \sim 10^4$~\hcc\ when the gas is mostly transparent to X-rays (i.e., shielding is negligible).
In Paper~II we have shown that X-ray shielding is important only for strong X-ray irradiation. Although a moderate X-ray background lowers the critical halo mass for star formation (Paper~I), it is the opposite in an intense X-ray background where heating by X-rays is the dominant feedback. For strong X-ray irradiation the gas in a halo can condense only if it is sufficiently massive ($M>M_{Jeans, IGM}$), therefore in this case the halo can be optically thick to X-rays. This is, in a sense, consistent with \citet{hummel2015} who pointed out the dependence of shielding on halo properties.
\subsection{Protostar Formation Recipes}
\label{app:clump}
\begin{figure*}
    \centering
	\includegraphics[width=0.38\textwidth]{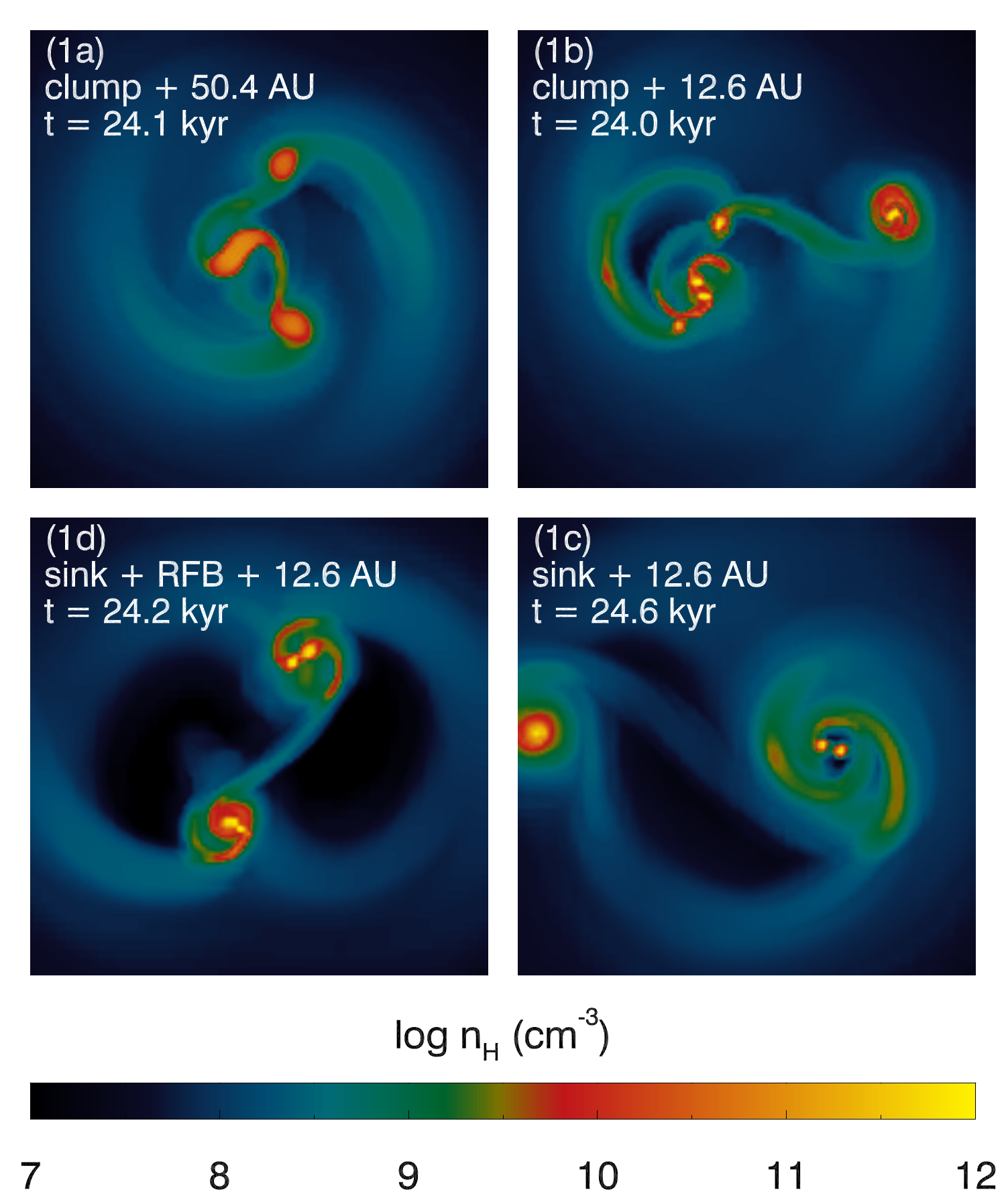}
	\includegraphics[width=0.48\textwidth]{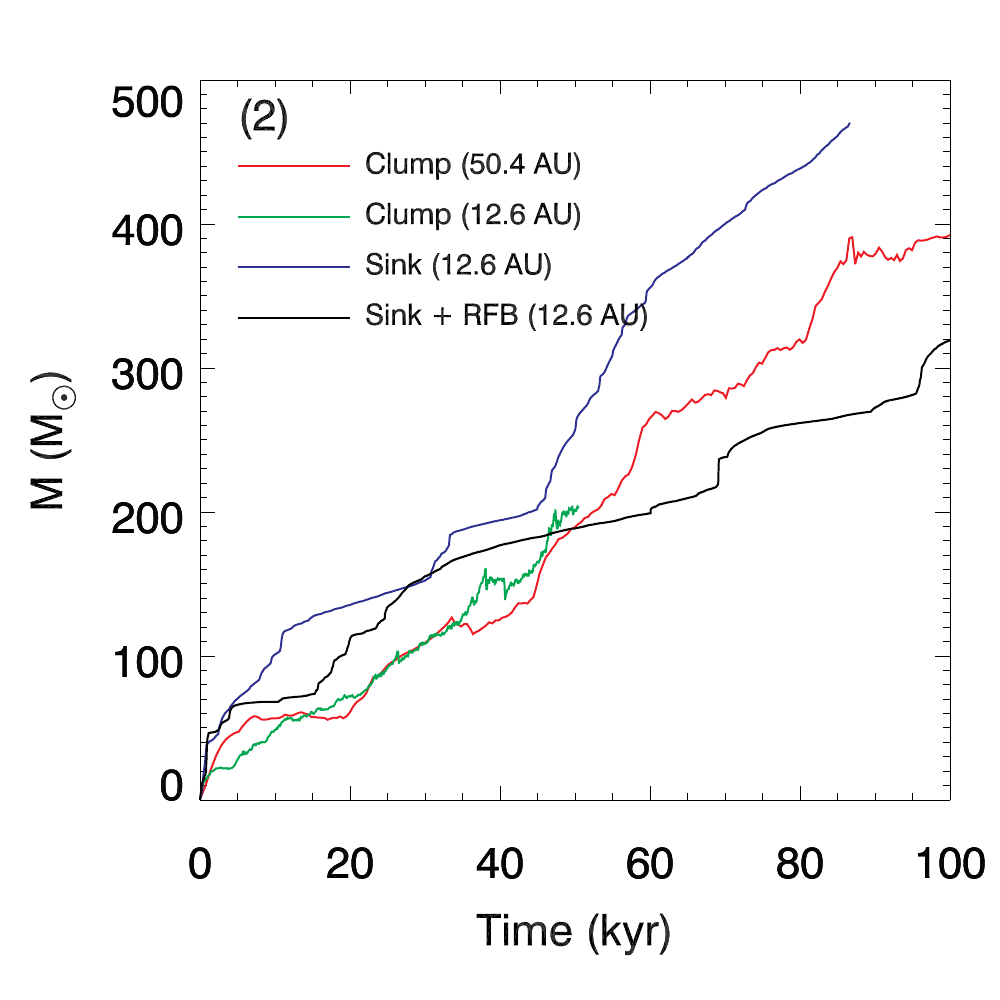}
    \caption{Panel (1a)-(1d): Slices of the hydrogen number density of Run~F with different star formation recipes. The figure shows the discs at $t \sim 24$~kyr since the formation of the first protostar. The four panels show the change with the improvement of the simulation from the top left panel in the clockwise direction. (1a): Low resolution with the clump recipe. This setting is similar to that of Paper~I and Paper~II. (1b): The spatial resolution is increased. (1c): Clumps are replaced by sink particles. (1d): RFB is added. This is the fiducial run. Panel (2): Evolution of the total mass in Pop~III stars for the same four protostar formation recipes as in the left panel.}
    \label{fig:nosink1}
\end{figure*}

Fig.~\ref{fig:nosink1} (1a-1d) shows snapshots of Run~F with different protostar formation recipes. Protostars in the low-resolution simulation (top left) show a similar evolution to the ones in Paper~I and Paper~II. With a higher spatial resolution (top right) clumps reach higher densities ($\sim 10^{13}$~\hcc) and fragmentation becomes more efficient creating more protostars. When sink particles are used, however, the number of stars is reduced due to sink mergers (bottom right). The change when including RFB is not prominent as seen in the bottom left panel.

In Fig.~\ref{fig:nosink1} (panel 2) we plot the total mass in Pop~III stars for the different recipes shown in the left panel. One notable feature is that the total mass of the clumps is independent of the spatial resolution (red and green). The total mass of protostars tracked by sink particles (blue) is higher than the one for the clumps, suggesting that the stars in the two methods are not identical. These three cases all show a linear growth of mass due to the lack of RFB.

\subsection{Sink Formation Threshold and Radius}
\label{app:sink}
In our simulations we want to resolve the Jeans length with at least $N_J>4$ cells to avoid numerical fragmentation \citep{truelove1997}. Therefore we impose a refinement criterion $\Delta x = \lambda_J/N_J$, with $N_J=16$. However, when $\Delta x=\Delta x_{min}$ at the maximum refinement level, the Jeans length is no longer resolved with $N_J$ cells if the density becomes too large. To avoid this we introduce sink particles that limit the increase of the density within their radii $r_{sink}=N_{Sink} \Delta x_{min}$ (note that inside the sinks the grid is always refined to the maximum level). A sink is formed when the peak density in a gas clump (identified by a clump finder) is above the critical density, $n_{sink}$, such that
\begin{equation}
    N_{Sink} \Delta x_{min} = \lambda_J=\left( \gamma \frac{k_{B}T}{\mu m_{H}} \right)^{1/2} \left( \frac{1}{G m_{H} n_{sink}} \right)^{1/2},
\end{equation}
where $N_{Sink}\le N_J$ is an integer. Solving for $n_{sink}$ gives,
\begin{equation}
    n_{sink} \sim 1.2 \times 10^{12}~{\rm cm}^{-3}
    \left( \frac{4}{N_{Sink}} \right)^2 
    \frac{(T/{1000~\rm K})}{(\Delta x_{min}/{12.6~\rm AU})^2}.
    \label{eq:nsink}
\end{equation}
We adopt a value $n_{sink}=10^{12}$~\cc\ in all our simulations, roughly corresponding to $N_{Sink}=4$, assuming $T=1000$~K (i.e., we resolve the Jeans length with 4 cells within the sink radius. As discussed in Appendix of Paper~I our chemistry and cooling implementation are valid up to $n \sim 10^{14}$~\cc\ and our choice of the density threshold is well below this limit.

\begin{figure}
    \centering
	\includegraphics[width=0.48\textwidth]{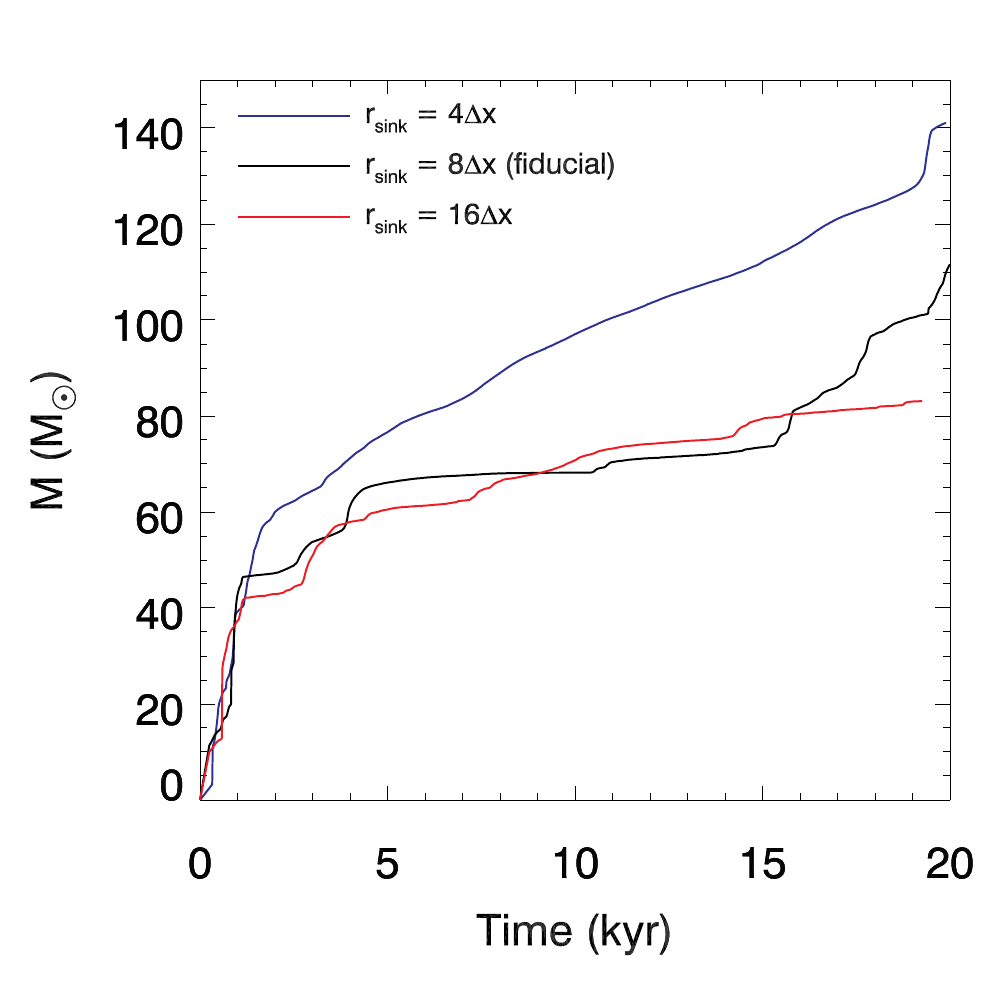}
    \caption{Sink mass with different sink radii. The fiducial value is $8 \Delta x$ where $\Delta x = 12.6$~AU. }
    \label{fig:rsink}
\end{figure}

The choice for the value of the sink radius is independent of the density threshold $n_{sink}$, but it makes sense to adopt values that are comparable or multiples of the Jeans length: in our case we explore the range $4\le N_{Sink}\le 16$. As illustrated in Fig.~\ref{fig:rsink} adopting $N_{Sink}=4$ with our sub-sink radiation transfer recipe in \ref{app:inj}, reduces significantly the RFB efficiency with respect to using $N_{Sink}=8,16$, producing a total mass in Pop~III stars comparable to the no-FB case. The total mass in Pop~III stars for $N_{Sink}=8$ and $N_{Sink}=16$ are similar, but start diverging after $t \sim 15$~kyr as more efficient feedback in the latter evaporates the gas disc. In our fiducial simulations we adopt $N_{Sink}=8$, but clearly $N_{Sink}$ is a crucial parameter that determines the strength of RFB and the spatial resolution of the simulations.

\subsection{Ionizing and LW Photon Emission from Sink Particles}
\label{app:inj}

Pop~III protostars form at the centre of sink particles, however the disc structure and dynamics inside the sink are poorly resolved. If we make a simple estimate of the size $R_S$ (Str\"{o}mgren radius) of the \HII region produced by a massive star at the centre of a sink in which there is a uniform density $n$, we must require that $R_S$ is greater than the sink radius for the ionizing radiation to escape the sink:
\begin{equation}
    \left( \frac{3Q}{4\pi n^2 \alpha} \right)^{1/3} \ge r_{sink},
\end{equation}
where $\alpha \approx 2 \times 10^{-10}T^{-3/4}~{\rm cm}^3~{\rm s}^{-1}$ is the recombination rate. Assuming $T=10^4$~K and an ionizing photon emission rate $Q=10^{50}~{\rm s}^{-1}$, we get $n \lesssim 10^{8}~{\rm cm}^{-3}$. This is several orders of magnitude smaller than the $n_{sink}$, hence the ionizing radiation should remain trapped inside the sink. 

However, the gas inside the sink does not have a constant density. Along the disc plane, the density is comparable to $n_{sink}$, however above and below the disc plane the density is much lower and, if the disc is well resolved, around the protostar the disc is thinner due to the star's gravity. For this reason, radiation can escape along bipolar \HII regions above and below the disc, while being trapped along the disc direction. Because the gas structure is poorly resolved inside the sink we do not inject radiation from the star at the centre of the sink or uniformly inside the sink particle, as it likely would underestimate RFB. Instead, we inject the radiation from the star at the surface of the sink sphere, with flux in the radial direction. We adopt a sub-sink radiative transfer prescription to take into account the ionizing and LW radiation absorbed by the gas inside the sink.
\begin{figure*}
    \centering
	\includegraphics[width=0.48\textwidth]{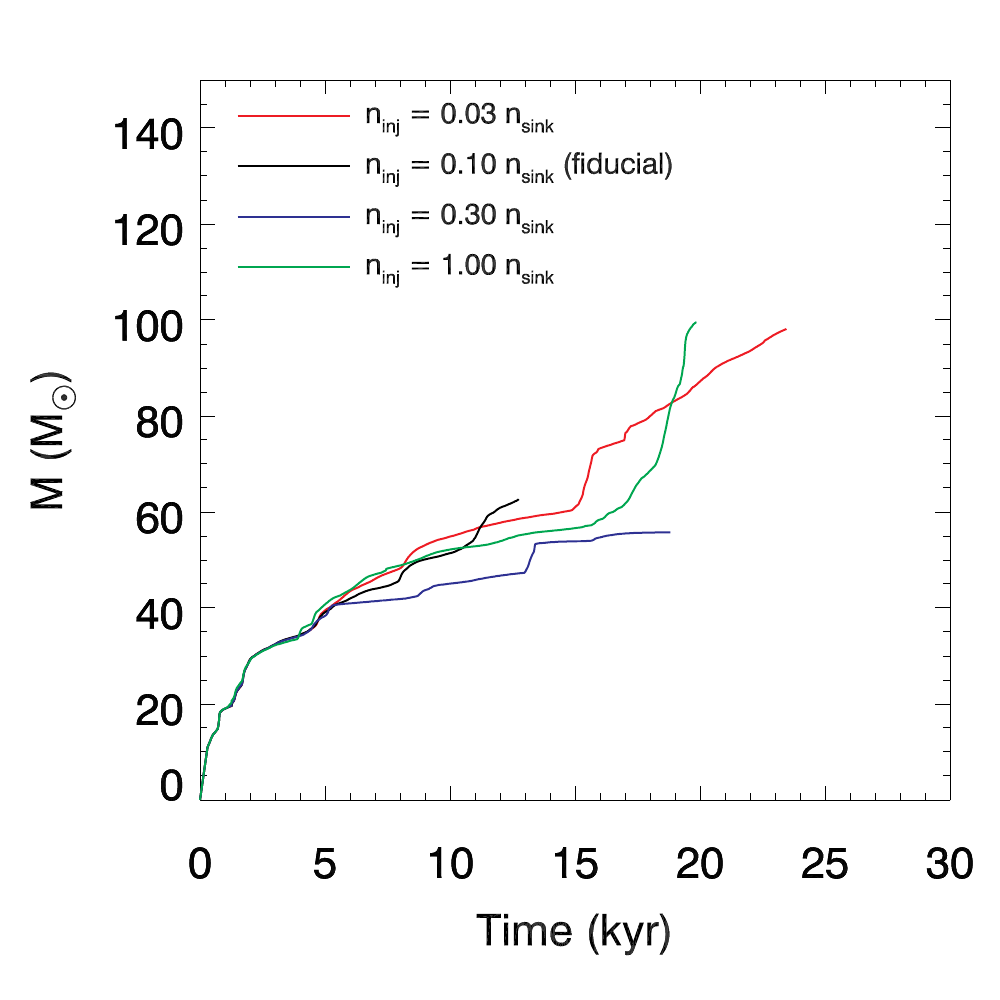}
	\includegraphics[width=0.48\textwidth]{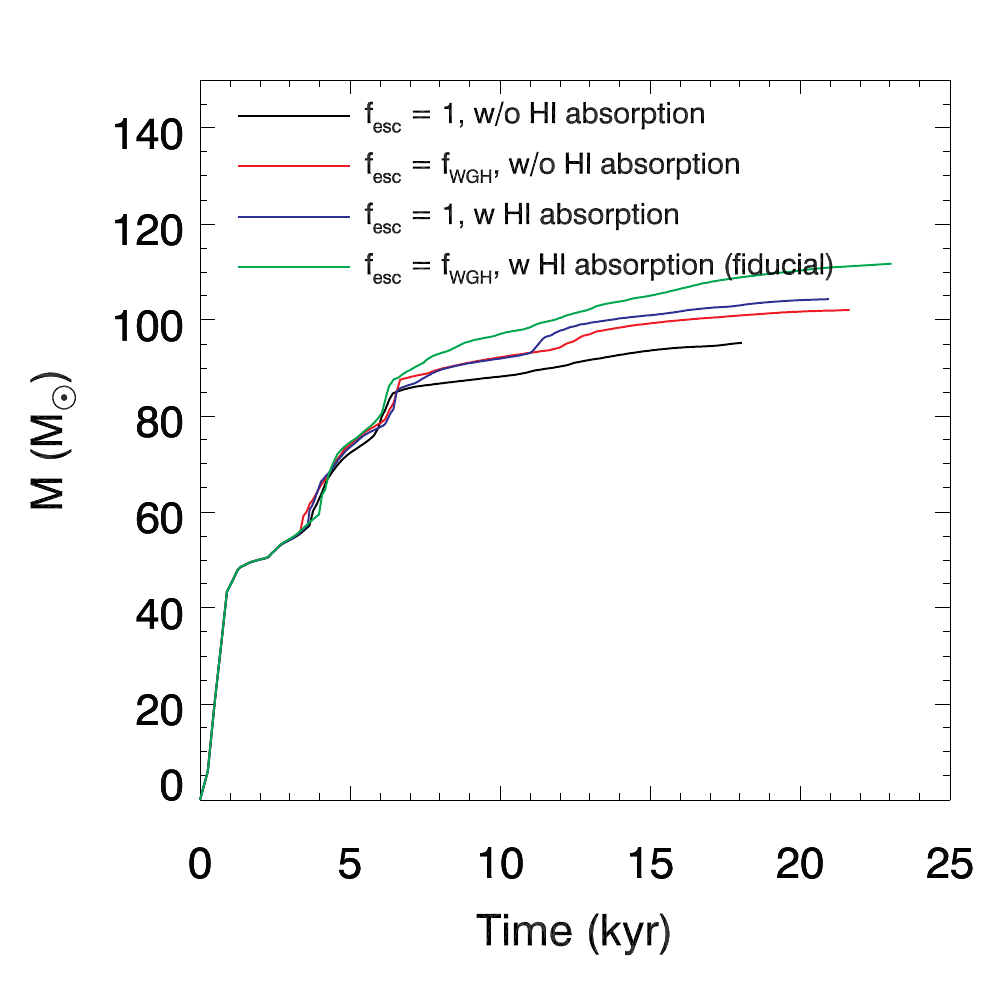}
    \caption{Left: Test runs with different $n_{inj}$. $0.1 n_{sink}$ is the fiducial value. Right: Sink mass affected by LW-related physics. The black line shows mass without a sub-sink recipe (i.e. $f_{esc}=1$) and \HI absorption. Adding sub-sink recipe (i.e. $f_{esc}=f_{WGH}$, equation~\ref{eq:esc}, red line) and \HI absorption (blue line) alone reduce RFB and thereby increasing stars' mass. Considering both (green line) increases the mass further. We assume that LW radiation is absorbed by \hm\ gas in all cases.}
    \label{fig:injection}
\end{figure*}

Regarding H and He ionizing radiation we fully absorb the radiation for rays where the gas density at the sink radius is higher than an
arbitrary threshold $n_{inj} = 0.1 n_{sink} = 10^{11}~{\rm cm}^{-3}$, since the radiation would be trapped anyway and we want to avoid photo-evaporation of the disc due to numerical effects. Test results with different density thresholds are shown in Fig.~\ref{fig:injection} (left). The opening angle of the bipolar region where ionizing radiation escapes is not sensitive to the assumed $n_{inj}$ within the explored range, and the total mass in Pop~III stars do not show any clear trend with the assumed value of $n_{inj}$. However, the values of $r_{sink}$ and $n_{inj}$ should be considered related parameters because if $r_{sink}$ is small, the gas density at the sink surface may never drop below $n_{inj}$ and radiation would remain trapped.

For the LW radiation, in addition to a density threshold, we calculate the column density of the \hm\ gas along rays from the centre of the sink to the sink surface and we estimate the fraction of LW photons escaping the sink in each direction due to \hm\ self-shielding. In Fig.~\ref{fig:injection} (right) we show how the total mass in sinks changes when including \hm\ self-shielding and absorption by \HI Lyman series lines (Section~\ref{sec:transfer}). The escape fraction of LW photons due to \hm\ self-shielding alone (red line) reduces the amount of LW photons escaping from the sink surface, thereby weakening RFB and increasing the total mass in Pop~III stars. The feedback effect also becomes weaker by a comparable amount due to shielding by \HI resonant lines alone (blue line). The total mass in Pop~III stars increases further when considering both sources of opacity (green line).

\subsection{Resolution Study}
\label{app:res}
\begin{figure}
    \centering
	\includegraphics[width=0.48\textwidth]{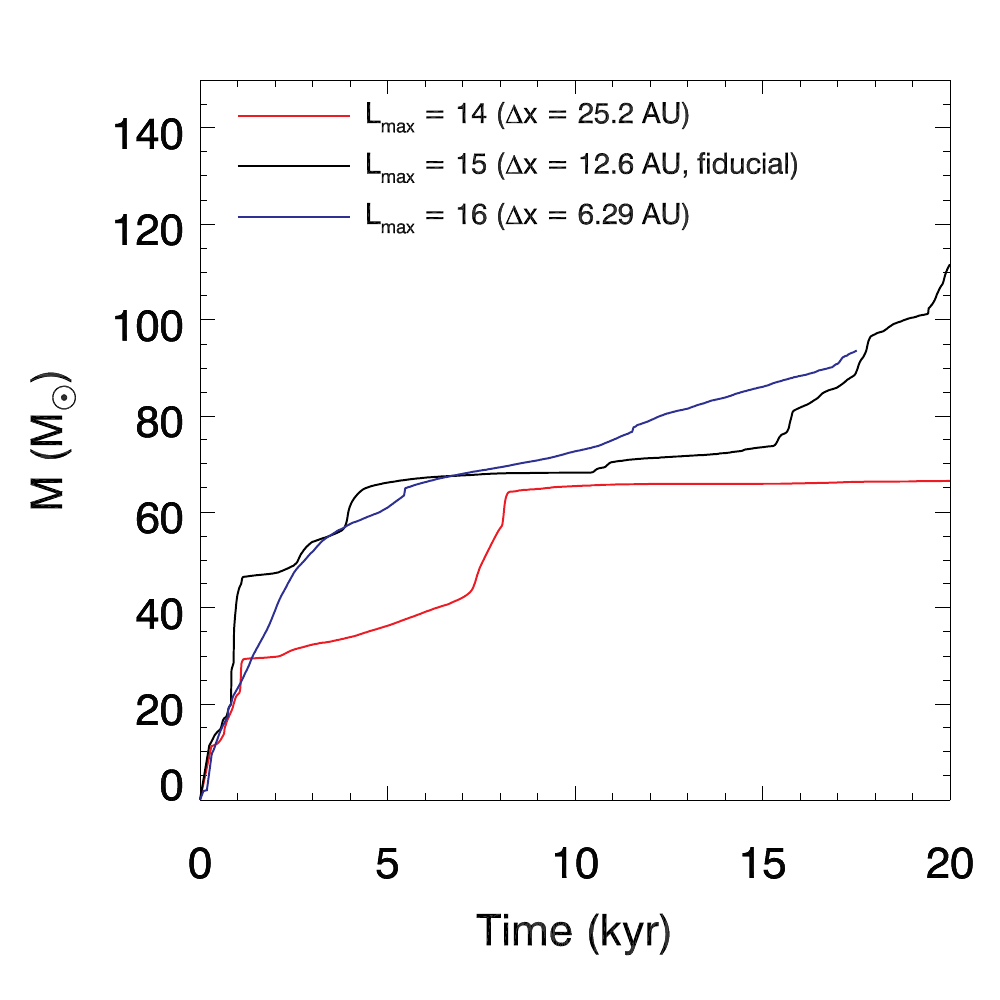}
    \caption{Resolution study. We perform Run~F with the maximum AMR level 14, 15 (fiducial), and 16. We also provide the corresponding cell sizes. Following equation~(\ref{eq:nsink}) we choose $n_{sink} = 2.5 \times 10^{11}, 1.0 \times 10^{12}$, and $4.0 \times 10^{12}$~\hcc.}
    \label{fig:res}
\end{figure}

Fig.~\ref{fig:res} shows the total mass in Run~F at different spatial resolutions $\Delta x_{min}$. The fiducial and high-resolution simulations show similar masses in the first $\sim 17$~kyr while the mass in the low-resolution case is lower.

\subsection{Reduced Speed of Light Approximation}
\label{app:rsla}

Ignoring recombination, the speed of the ionization front is
$v_I = Q/(4 \pi d^2 n)$,
where $d$ is the distance from the source. By normalizing the parameters with typical values relevant to this work, at large scales we get:
\begin{equation}
        v_I= 2.79 \times 10^{-4} c~\left( \frac{Q_{48}}{d_{0.01pc}^2 n_7}\right).
\end{equation}
Here, $Q_{48}$ is the luminosity of the source in units $10^{48}$~photon~s$^{-1}$, $d_{0.01pc}$ is the radius of the ionization front in units of $0.01$~pc, and $n_7$ is the gas number density in units of $10^7~{\rm cm}^{-3}$. Therefore at large distances, assuming $Q_{48}\sim 1-10$, $v_I$ is about  a factor of $3-30$ times slower than our fiducial value for the reduced speed of light $c_{red}=10^{-3}c$.
We expect the I-front propagation speed to be similar near the sink particles because the distance decreases but the gas density increases roughly as $\sim r^{-2}$ near the sink \citep{larson1969,penston1969}, therefore the ionization front speed only depends on the luminosity of the source. With our sink radius of $\sim 100$~AU ($d_{0.01pc} \sim 5 \times 10^{-2}$) and photon injection density threshold $10^{11}$~\hcc\ ($n_7 = 10^4$), the ionization front speed is $v_I \sim 10^{-5} Q_{48}c$. Therefore even for $Q_{48} \sim 10 - 100$, only relevant when sinks have high mass and accretion rates, $v_I \sim 10^{-4} - 10^{-3}c$. All the estimates above for $v_I$ are only an upper limit for the speed, as recombination slows down the ionization front propagation speed. Indeed we have tested different values of $c_{red}$ from $10^{-1}c$ to $10^{-4}c$, and did not observe significant effects on the simulations results. Interestingly the computational cost of the simulations is not reduced significantly when assuming values of $c_{red}$ lower than $10^{-3}c$, hence we adopted the most conservative $c_{red}=10^{-3}c$.

\section{Relative Importance of EUV vs FUV Feedback}
\label{app:UV}
\begin{figure}
    \centering
	\includegraphics[width=0.48\textwidth]{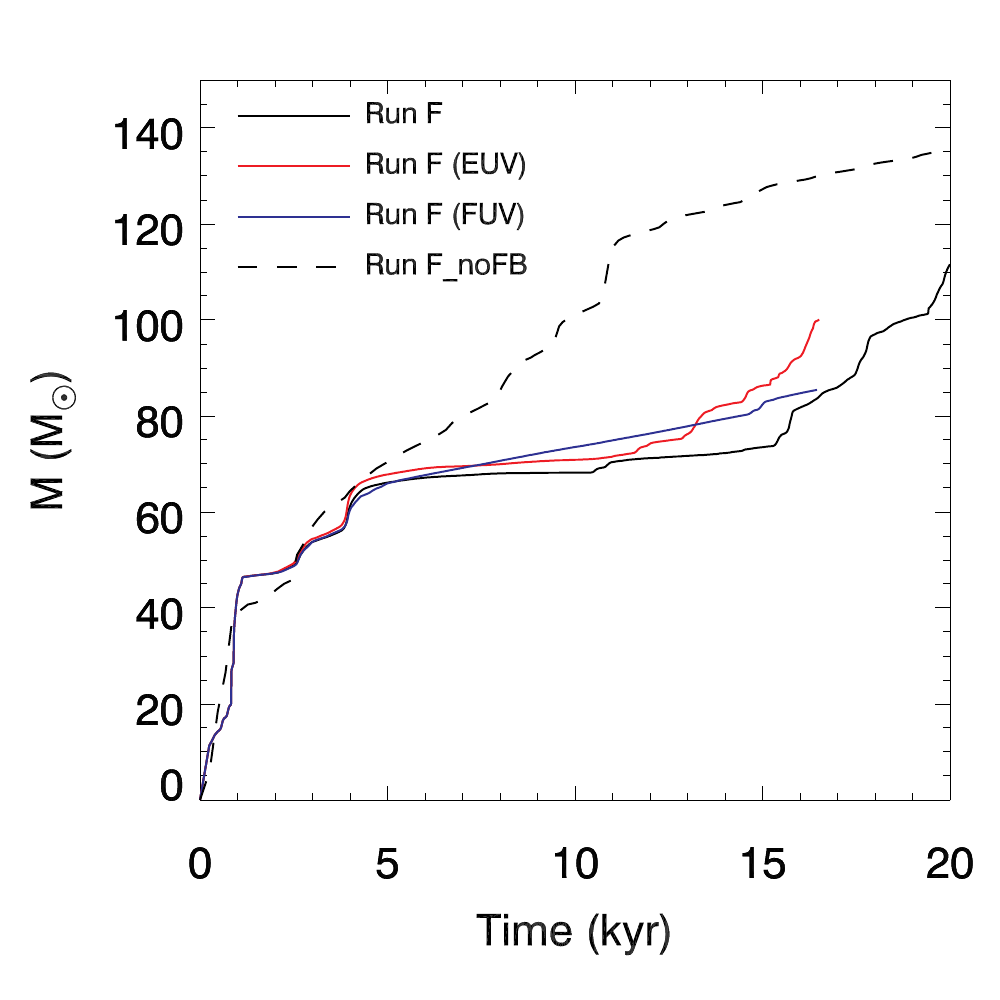}
    \caption{Total mass in sinks in Run~F affected by EUV (red), FUV (blue) and both (black solid). For comparison noFB case is shown in a dashed line.}
    \label{fig:UV}
\end{figure}

In Fig.~\ref{fig:UV} we plot the total mass in sinks as a function of time in runs with different UV feedback. We find FUV (LW, red line) alone can efficiently reduce the mass unlike in \citet{hosokawa2016}. EUV-only simulation shows a similar degree of mass reduction. The mass is reduced further with the combination of two UV feedback but this further reduction seems weaker than individual feedback.

\section{Sink Merger}
\label{app:merger}
\begin{figure}
    \centering
	\includegraphics[width=0.48\textwidth]{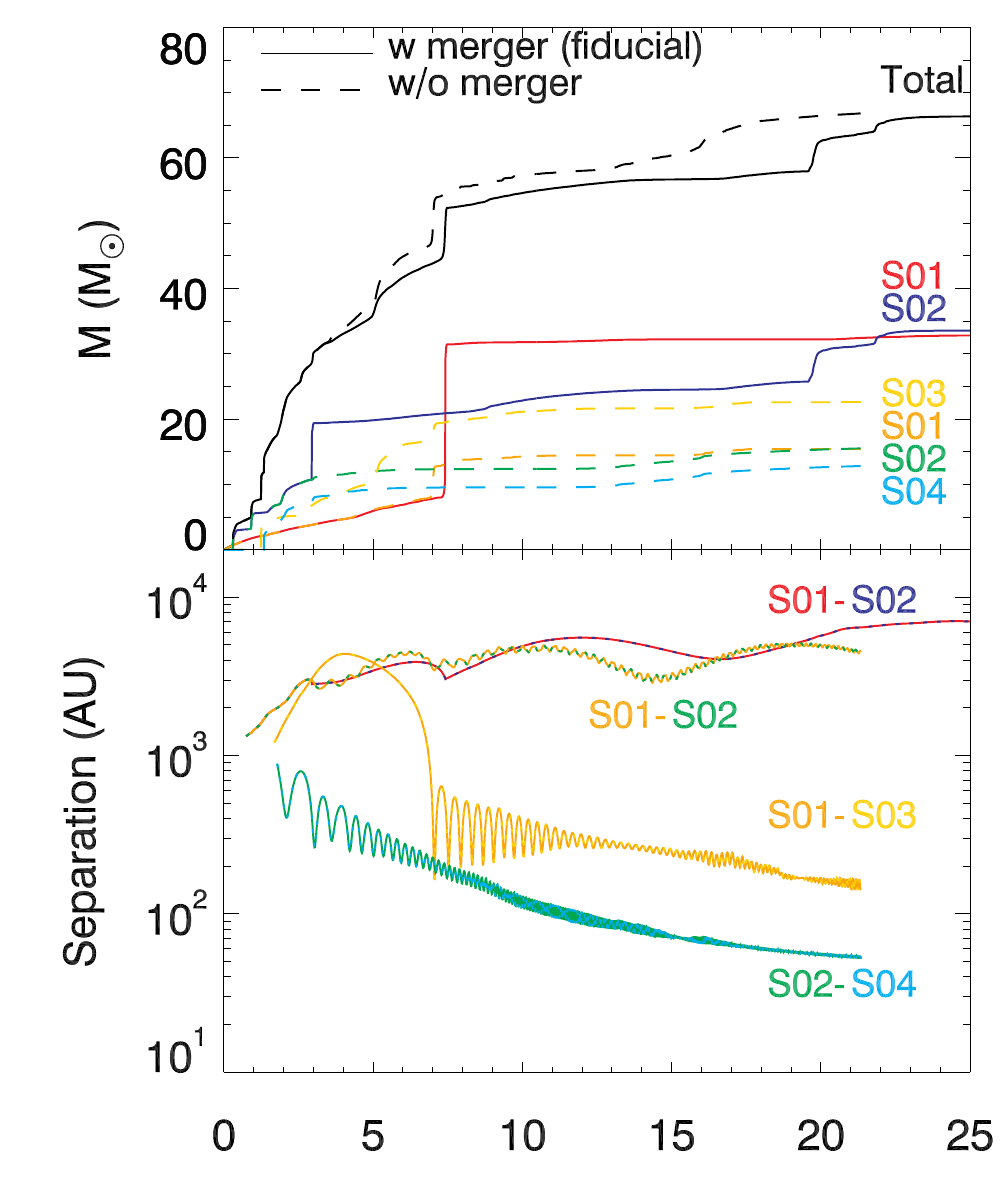}
    \caption{Top: Masses in Run~E with (solid lines) and without (dashed lines) sink mergers. The multiplicities are 2 and 4, respectively. Individual sink particles are shown in different colors. Bottom: Binary separations. In the simulation without mergers the stars form a hierarchical binary.}
    \label{fig:merger}
\end{figure}

In Fig.~\ref{fig:merger} we plot the mass in Run~E with and without sink merger (solid and dashed). In the fiducial run, 4 stars form and merge to leave two stars. When they merge, they are gravitationally bound to their companions. In the test run, these stars actually orbit their companions in close orbits with separations $\sim 40$ and $100$~AU. The orbits do not expand but shrink with time possibly due to gravitational torque by the gas disc \citep{chon2019} and they circularise with time. RFB in the test run is slightly weaker than the fiducial case because the luminosity of a protostar is not a linear function of mass in the model \citep{hosokawa2009,hosokawa2010}. Therefore the feedback strength may differ even if the total mass is the same. Even with this difference, the mass variation is insignificant. The total masses of the stars at $\sim 22$~kyr in the fiducial and test runs are $63.7$ and $66.8~\msun$, respectively. We expect the mass of stars is not sensitive to the sink merger prescription.


\bsp	
\label{lastpage}
\end{document}